\documentclass[conf]{new-aiaa}
\usepackage[utf8]{inputenc}

\usepackage{graphicx}
\usepackage{amsmath}
\usepackage[version=4]{mhchem}
\usepackage{siunitx}
\usepackage{longtable,tabularx}
\usepackage{comment}
\usepackage{xcolor}
\usepackage{booktabs}
\usepackage{algorithm}
\usepackage{algpseudocode}

\newcommand{\pp}[2]{\frac{\partial #1}{\partial #2}}

\newcommand{\ol}[1]{\overline{#1}}

\newcommand{\bx}{\mathbf{x}}
\newcommand{\bu}{\mathbf{u}}
\newcommand{\bv}{\mathbf{v}}
\newcommand{\bq}{\mathbf{q}}

\newcommand{\bU}{\mathbf{U}}
\newcommand{\bF}{\mathbf{F}}
\newcommand{\Kn}{\mathrm{Kn}}
\newcommand{\Mtr}{M_\mathrm{train}}
\newcommand{\relErr}{{\epsilon_\mathrm{rel}}}

\newcommand\solidrule[1][0.5cm]{\rule[0.5ex]{#1}{.4pt}}
\newcommand\dashedrule{\mbox{%
\solidrule[1mm]\hspace{0.65mm}\solidrule[1mm]\hspace{0.65mm}\solidrule[1mm]\hspace{0.65mm}\solidrule[1mm]}}

\title{Deep Learning Closure of the Navier--Stokes Equations for Transition-Continuum Flows
\footnotetext{Aspects of this work were presented as Paper 2023-1796 at the 2023 AIAA SciTech Forum, National Harbor, MD, January 23--27, 2023, and Paper 2022-1703 at the 2022 AIAA SciTech Forum, San Diego, CA, January 3--7, 2022.}
}

\author{Ashish S. Nair\footnote{Graduate Student, Department of Aerospace and Mechanical Engineering}}
\affil{University of Notre Dame, Notre Dame, Indiana 46556, USA}

\author{Justin Sirignano\footnote{Associate Professor, Mathematical Institute}}
\affil{University of Oxford, Oxford OX2 6GG, UK}

\author{Marco Panesi\footnote{Professor, Department of Aerospace Engineering;  Director, Center for Hypersonic and Entry Systems Studies. Associate Fellow AIAA.}}
\affil{University of Illinois at Urbana--Champaign, Urbana, Illinois 61801, USA}

\author{Jonathan F. MacArt\footnote{Assistant Professor, Department of Aerospace and Mechanical Engineering; jmacart@nd.edu. Member AIAA.}}
\affil{University of Notre Dame, Notre Dame, Indiana 46556, USA}

\begin{document}

\maketitle

\begin{abstract}
  The predictive accuracy of the Navier–Stokes equations is known to degrade at the limits of the continuum assumption, thereby necessitating expensive and approximate solutions to the Boltzmann equation. While tractable in one spatial dimension, their high dimensionality increases the computational challenge of multi-dimensional, multi-physical Boltzmann calculations. It is therefore desirable to augment the Navier–Stokes equations for accuracy under these conditions. We present an application of a deep learning method to extend the validity of the Navier–Stokes equations to transition-continuum flows. The technique encodes the missing physics via a neural network, which is trained to reduce the error between the Navier--Stokes and Boltzmann solutions. While standard DL methods can be considered \emph{ad hoc} due to the absence of underlying physical laws, at least in the sense that the systems are not governed by known partial differential equations, the DL framework leverages the \emph{a priori} known Boltzmann physics while ensuring that the trained model is consistent with the Navier–Stokes equations. The online training procedure solves adjoint equations, constructed using algorithmic differentiation, which efficiently provide the gradient of the loss function with respect to the learnable parameters.  The model is trained and applied to predict stationary, one-dimensional shock thickness in low-pressure argon.

\end{abstract}

\section{Nomenclature}

{\renewcommand\arraystretch{1.0}
\noindent\begin{longtable*}{@{}l @{\quad=\quad} l@{}}
$\mathbf{U}$  & dependent variables \\
$\hat{\mathbf{u}}$  & adjoint variables \\
$\theta$ & model parameters \\
$f(\mathbf{U}_x,\theta)$  & neural network closure model \\
\end{longtable*}}

\section{Introduction}
\lettrine{C}{omputationally} efficient and accurate models for flows in translational nonequilibrium are essential for designing, operating, and developing hypersonic flight vehicles and their advanced propulsion systems.  The Knudsen number ($\Kn=\lambda/L$, where $\lambda$ is the molecular mean free path and $L$ is a characteristic length scale) indicates the flow regime, spanning continuum to free-molecule flow. Increasing $\Kn$ indicates an increasing departure from local thermodynamic equilibrium, which affects the applicability of various flow models~\cite{Paolucci2018}. For $\Kn \le 10^{-3}$, the continuum Navier--Stokes equations are an appropriate fluid model. However, existing computationally efficient hydrodynamic models, perturbative in nature~\cite{Paolucci2018}, become unreliable as $\Kn$ rises due to either low densities or regions of high local gradients (as in hypersonic flows), which cause deviations from the linear Newtonian and Fourier laws. The high-Knudsen-number regime can be further divided into the transition-continuum regime ($0.01 \le \Kn \le 1$) and the free-molecule flow regime ($\Kn \ge 1$).

From a theoretical standpoint, the kinetic description of interacting particles via the solution of the Boltzmann equation allows for the most accurate description of nonequilibrium hypersonic flows. 
Because of the difficulties associated with the solution of the Boltzmann equation, approximate physics-based methods---at different spatial and time scales---have been developed over the years by the scientific community. This section aims to provide a brief overview and a critical analysis of the approaches currently used in the literature to alleviate the computational challenges of solving the Boltzmann equation.

These methods can be classified into two broad categories: Direct Simulation Monte Carlo (DSMC) and extended-hydrodynamics methods. DSMC involves the simulation of collisions between many ``computational molecules''  that statistically describe the physical dynamics and interactions among fluid molecules~\cite{Bird1963,Bird1970}. While popular, DSMC remains challenging for  mixed  transition-continuum and continuum flows, though modern DSMC can solve complicated, multiphysical flows given sufficient computing power \cite{Boyd2007,Boyd2015}. 
The computational cost of DSMC  can be alleviated by extended-hydrodynamics methods, which use the fluid equations in conjunction with higher-order constitutive models for the viscous stress and heat flux terms to extend their validity beyond the continuum regime. Many of these methods use the Chapman--Enskog expansion, which uses $\Kn$-series approximations (of varying order) to the Boltzmann equation. Retaining the zero-order terms gives the Euler equations, first-order terms the Navier--Stokes equations, and second-order terms the Burnett equations~\cite{Chapman1970}. The Navier--Stokes equations are valid in the continuum limit ($\mathrm{Kn}\rightarrow 0$) of small mean free paths~\cite{Chapman1970,Ferziger1972MathTransport}. However, large perturbations of the velocity distribution from equilibrium render the continuum assumption invalid. Nonperturbative moment methods~\cite{Levermore96} mitigate the problem of the breakdown away from moderate nonequilibrium by predicting values for momentum and energy fluxes that are consistent with the nonnegativity of the particle density. These methods, however, are storage-intensive and computationally impractical due to the large number of independent variables required to evaluate integral terms. Several variants of the Burnett equations have been proposed including the BGK--Burnett (Bhatnagar–Gross–Krook) equations~\cite{Agarwal2001}, Woods equations~\cite{Woods1993}, Grad's 13-moment equations~\cite{Grad1949,Grad1952}, regularized Burnett~\cite{Jin2001}, and regularized Grad's 13-moment equations~\cite{RegBurett2003}. A major drawback of most is their mathematical ill-posedness in their original formulations, which leads to physical and numerical instabilities. Thermodynamic consistency is also not an intrinsic property of these equations; this limits their general applicability, despite their computational cost being comparable to the Navier--Stokes equations. The advantages of DSMC's detailed kinetic information and the continuum-like methods' computational tractability can be obtained by hybrid schemes~\cite{Hash96Hybrid,Aktas2002Hybrid,Panesi2011Hybrid,stephani2012consistent, stephani2013non, schwartzentruber2006hybrid, schwartzentruber2008hybrid, schwartzentruber2008hybrid2, schwartzentruber2007modular, burt2009hybrid}, which solve the Navier--Stokes equations in continuum regions of the flow and employ the DSMC technique elsewhere. Although these methods are attractive for multiscale flows, identifying the continuum and DSMC (molecular) regions is often done from the continuum perspective, numerically interfacing these regions is challenging, and the speedup obtained is often insufficient to render it an attractive alternative to pure DSMC. Deterministic numerical methods that directly solve the Boltzmann equation~\cite{Munafo2012Direct} have been tested for simple flow configurations, though their extension to practical problems is yet to be explored.

This paper describes  the construction of a physics-based reduced-order model, able to capture the fundamental physical processes that occur in hypersonic flows, yet computationally efficient, by leveraging fundamental physics, computational science, and applied mathematics. The proposed model compensates for the deficiencies of the Navier--Stokes model in the transition-continuum regime by introducing additional terms into the continuum equations. By estimating the parameters of these terms (we use neural networks) while converging the Navier--Stokes solutions to Boltzmann solutions, the model attempts to encode the unrepresented physics that are missing from the continuum description of the gas. We also propose algebraic constraints to the model outputs to ensure consistency with the second law of thermodynamics.

Applications of deep learning (DL) for closure modeling are widespread in incompressible fluid mechanics~\cite{Brenner2019,duraisamy2019turbulence,duraisamy2021perspectives,karniadakis2021physics,Kochkov2021} but are much less common for compressible flows and hypersonics. Most previously employed machine learning methods for flows attempt to estimate unclosed terms (e.g., the Reynolds stress or the subgrid stress) by minimizing the direct mismatch between these terms and ``trusted'' data, which are often obtained from highly resolved direct numerical simulations (DNS)~\cite{ling2016machine, wu2018physics, beetham2021sparse}. This approach, which we call \emph{a priori} training, decouples the parameter estimation (optimization) step from the solution of the governing partial differential equations (PDEs), which is mathematically inconsistent due to the noncommutativity of nonlinear operations~\cite{Sirignano2020}. In our view, this approach has two primary deficiencies for compressible flows. First, the nonlinear coupling of dependent variables inherent to compressible flows intensifies the inconsistency of \emph{a priori} training. Second, and more importantly, when the discrepancy is due to a fundamental deficiency in the PDE model---for example, for transition-continuum flows, approximations to the viscous stress and heat flux from kinetic theory---the functional form of the discrepancy is often unknown; thus, even defining the loss function for the direct mismatch is challenging.

Our deep learning approach is conceptually simple yet avoids these challenges. Rather than decouple the optimization step from the PDE solution, we use an \emph{embedded optimization} method that substitutes the untrained model into the PDE and estimates its parameters \emph{in situ}~\cite{Sirignano2020, holland2019towards, duraisamy2021perspectives}. The model thus respects its effects on the nonlinearly coupled PDE solution. Additionally, this approach can directly compute the loss function from the dependent variables (mass, momentum, and energy) or derived quantities (e.g., temperature) rather than from higher-order unclosed terms appearing in the PDEs. This allows us to obtain target data from Boltzmann solutions and, in theory, enables straightforward incorporation of experimental data. We present an embedded optimization approach that optimizes over the PDE solution by solving the adjoint Navier--Stokes equations, which are obtained using algorithmic differentiation (AD) of the forward equations and are solved efficiently using graphics processing unit (GPU) acceleration. Similar methods have been successfully applied to turbulence modeling for incompressible large-eddy simulation (LES)~\cite{Sirignano2020,MacArt2021,sirignano2022BluffBodies}, Reynolds-averaged Navier--Stokes (RANS) simulation~\cite{sirignano2021pdeconstrained, holland2019towards} and sensitivity analysis of chaotic dynamical systems~\cite{blonigan2018multiple, ni2019sensitivity}.  Given the code-intrusive nature of adjoint-based optimization, many efforts to avoid this difficulty have used less-complete couplings of the optimization step and the PDE solution~\cite{taghizadeh2020turbulence, zhao2020rans, liu2021iterative}, though these methods typically require trusted data for the exact term being modeled, which is not always possible.

This work presents an example application to the prediction of the shock structure in rarefied argon flows, compares the augmented Navier--Stokes predictions to the unmodified predictions, assesses the Mach number interpolation and extrapolation capability of the DL model, and provides a physical interpretation of the modifications to the continuum transport terms.
 The governing equations and transport models from kinetic theory are briefly reviewed in Section~\ref{sec:eqs}. The embedded deep learning framework and constraints to enforce the second law of thermodynamics are introduced in Section~\ref{sec:ML}. An application to transition-continuum argon shock predictions is presented in Section~\ref{sec:results}. A summary and discussion of future work are provided in Section~\ref{sec:conclusion}.

\section{Continuum Equations and Closure Models} \label{sec:eqs}

Let $g(t,\bx,\bv)$ be the solution to the Boltzmann equation (the distribution function) for $t\in[0,\infty)$ and $\bx,\bv\in\mathbb{R}^3$, and define the mass density  $\rho(t,\bx) = \langle m, g(t,\bx, \cdot) \rangle$ and fluid-dynamic momentum $\rho\bu(t,\bx) =  \langle m g(t,\bx,\cdot), \bv \rangle$, where  $m$ is the mass of a particle, and integrals of the distribution function over the velocity space are obtained from $\langle h(\cdot), g(t,\bx, \cdot) \rangle = \int_{\mathbb{R}^3} h(\bv) g(t,\bx, \bv) d\bv$. In this manner, all continuum quantities of interest $\bU=\{\rho,\bu,p,T,\dotsc\}$ may be obtained, where $p$ is the pressure, and $T$ is the temperature, among other possibilities.

The compressible Navier--Stokes equations for a single-component gas, as derived from the Boltzmann equation, are
\begin{align}
    \pp{\rho}{t} + \nabla\cdot(\rho\bu) &= 0 \notag \\
    \pp{\rho\bu}{t} + \nabla\cdot\Big[ \rho\bu\otimes\bu -
        \tau(\bu, \bv, g)\Big] &= 0 \label{eq:NS} \\
    \pp{\rho E}{t} + \nabla\cdot\Big[\rho E\bu - \tau(\bu, \bv, g) : \bu + \mathbf{q}(\bu, \bv, g)\Big] &= 0, \notag
\end{align}
where $E = e + \bu^\top\bu/2$ is the total energy and $e$ is the internal energy. The thermal momentum flux
\begin{equation}
  \tau(\bu, \bv, g) = \Big\langle m\bv\otimes\bv - \rho\bu\otimes\bu,g(t,\bx,\cdot) \Big\rangle 
  \label{eq:tau_unclosed}
\end{equation}
and the heat flux {$\bq(\bu, \bv, g)$} require integration over the Boltzmann phase space; thus, they are unclosed under the continuum variables. Closure of \eqref{eq:tau_unclosed} is commonly achieved by assuming proportionality to the strain-rate tensor and defining the pressure, yielding $\tau = \sigma - p\mathbf{I}$, where $\sigma = \mu ( \nabla \bu + \nabla \bu^{\top} ) - \frac{2}{3} \mu (\nabla \cdot \bu ) \mathbf{I}$, $\mu$ is the viscosity, and the pressure satisfies the ideal-gas equation of state $p = (\gamma-1)\left(\rho E - \rho\bu^\top\bu/2\right)$, where $\gamma$ is the (constant) ratio of specific heats. Similarly, the heat flux is typically closed using $\bq=-\lambda \nabla T$, where $\lambda$ is the thermal conductivity, $T=p/(\rho R)$, and $R$ is the specific gas constant. However, these closures---and the use of the Navier--Stokes equations more generally---require the continuum approximation to hold, which can inaccurately represent the physics of nonequilibrium flows in the transition-continuum regime.

\subsection{Neural-Network Closures}

We introduce three approaches for the closure of \eqref{eq:NS} using neural networks. First, the \textbf{transport coefficients} $\mu$ and $\lambda$ may be \textbf{corrected} using neural networks (Approach A),
\begin{align}
  \begin{split}
    \mu(\textbf{U};\theta) &= \mu_0(1+f_1(\textbf{U}_x ;\theta)), \\ 
    \lambda(\textbf{U};\theta) &= \lambda_0(1+f_2(\textbf{U}_x;\theta)),
    \end{split}
  \label{eq:mu_aug_model}
\end{align}
where $f_1(\textbf{U}_x;\theta)$ and $f_2(\textbf{U}_x;\theta)$ are the outputs of a neural network $f(\textbf{U}_x;\theta)$, which takes the point-wise derivatives of \textbf{U} as its input and has parameters $\theta$. The resulting  viscous stress and heat flux are
\begin{align}
  \begin{split}
    \sigma(\textbf{U}; \theta) &= \mu(\textbf{U}; \theta) \left(\ \left(\nabla\bu + (\nabla\bu)^\top\right) - \frac{2}{3} (\nabla\cdot\bu)\mathbf{I} \right) \quad \text{and} \\
    \bq(\textbf{U}; \theta) &= -\lambda(\textbf{U}; \theta) \nabla T.
    \end{split}
  \label{eq:mu_aug_model_closure}
\end{align}
Second,  the continuum viscous-stress and heat-flux models may be \textbf{augmented} by linearly superimposing neural networks (Approach B).  This results in augmented viscous-stress and heat-flux terms,
\begin{align}
  \begin{split}
    \sigma(\textbf{U}; \theta) &= \mu \left(\ \left(\nabla\bu + (\nabla\bu)^\top\right) - \frac{2}{3} (\nabla\cdot\bu)\mathbf{I} \right) - f_1(\textbf{U}_x ;\theta) \\ 
    \bq(\textbf{U}; \theta) &= -\lambda \nabla T - f_2(\textbf{U}_x;\theta),
  \end{split}
  \label{eq:aug_model}
\end{align}
where $f_1(\textbf{U}_x;\theta)$ and $f_2(\textbf{U}_x;\theta)$ are  again neural-network outputs, though they need not be the same networks as in \eqref{eq:mu_aug_model}.
Finally, the continuum models ${\sigma}(\bU)$ and ${\mathbf{q}}(\bU)$  may be \textbf{directly modeled} using neural networks (Approach C),
\begin{equation}
  \sigma(\bU;\theta) =  f_1(\textbf{U}_x;\theta)\qquad
  \text{and} \qquad
  \bq(\bU;\theta) = f_2(\textbf{U}_x;\theta).
  \label{eq:replacement_model}
\end{equation}
We pursue only Approaches A and B here. These have more limited potential than Approach C, being constrained by the algebraic models for $\sigma$ and $\bq$, but are more straightforward to implement and analyze. Rather than directly model the transport terms, our goal is to select $\theta$ such that the new models account only for \emph{deviations} of the classical relations from kinetic theory. The neural network $f(\textbf{U}_x;\theta)$ structure and optimization are discussed further in Section~\ref{sec:ML}.

An important uncertainty in the physical modeling of the shock structure, especially at high Mach numbers, is the temperature dependence of the transport coefficients $\mu$ and $\lambda$. We use a power-law dependence of the baseline continuum transport coefficients on temperature,
\begin{equation} 
    \mu(T) = \mu_0 \left(  \frac{T}{T_0} \right)^s
    \qquad \mathrm{and} \qquad
    \lambda(T) = \lambda_0 \left(  \frac{T}{T_0}\right)^s,
    \label{eq:mu_phys}
\end{equation}
where $T_0=273.15\,\mathrm{K}$, is typically used for argon. Several previous efforts have aimed to adjust the value of $s$ to reproduce experimental shock profiles~\cite{Alsemeyer1976, Reese1995, Bird1970, Lumpkin1991}. Following these, we use $s=0.74$ to match the viscosity law exponent used in our DSMC simulations.

\subsection{Application to 1D Shocks}

We consider an application to a 1D stationary shock in argon at $p_\infty=6.667$~Pa and $T_\infty=300$~K. This is a common shock-structure problem for  extended hydrodynamics models~\cite{Muckenfuss1962,Lumpkin1991,Reese1995,Levermore1998,Cercignani1998,Myong2001,Xu2004} that has been extensively studied experimentally~\cite{Alsemeyer1976, LiznerAndHornig, Camac, Schmidt1969, Garen1974}. The flow has a transition-continuum Knudsen number $\Kn=0.2 \sim 0.3$ within the shock, which Navier--Stokes equations do not accurately model. A monatomic gas such as argon is generally chosen to study the shock structure due to the lack of vibrational or rotational nonequilibrium. We assume a constant ratio of specific heats $\gamma=5/3$ with gas constant $R=208.12\,\mathrm{kJ}/\mathrm{kg}\cdot\mathrm{K}$. The 1D stationary solution of \eqref{eq:NS} satisfies
\begin{equation}
  \pp{\bU}{t} = -\frac{\partial {\bf F}_c}{\partial x} + \frac{\partial {\bf F}_d}{\partial x} = 0,
  \label{eq:NS_1d}
\end{equation}
where  
\begin{equation*}
{\bf U} = \left[ \begin{array}{cc} \rho \\ \rho u \\ \rho E \end{array} \right], \qquad 
{\bf F}_c = \left[ \begin{array}{cc} \rho u \\ \rho u^2 + p \\ \rho u H \end{array} \right], \qquad
{\bf F}_d = \left[ \begin{array}{cc} 0 \\ \sigma(\textbf{U};\theta) \\ \sigma(\textbf{U};\theta) u - q(\textbf{U};\theta) \end{array} \right],
\end{equation*}
where $H = e + p/\rho + 1/2 u^2$ is the total enthalpy, and $\sigma(\bU;\theta)$ and $q(\bU;\theta)$ are the 1D versions of the viscous-stress tensor and heat-flux vector with  embedded neural networks. $\bU$ satisfies Dirichlet (supersonic freestream) boundary conditions at $x = - \infty$ and subsonic nonreflective characteristic boundary conditions at $x = + \infty$~\cite{poinsot1992boundary}.

The 1D equations are solved on a domain $L_x\in[-20\,\mathrm{mm},10\,\mathrm{mm}]$ with the shock positioned at $x=0$. Inviscid shock solutions provide the initial conditions and Dirichlet boundary conditions. The equations are discretized over $n_x=256$ uniformly spaced mesh cells with the Euler fluxes discretized using a characteristic-based scheme~\cite{roe1986characteristic}; all other terms are discretized using second-order centered differences. The equations are converged to their steady-state solution using a damped Newton method; details are provided in Section \ref{sec: Newton solver}

Target data are obtained from DSMC solutions of the Boltzmann equation at freestream Mach numbers $M_\infty=2,3,5,6,8,9$, and $10$. The DSMC simulations used a VHS collision model and were initialized using the same pre- and post-shock conditions as the Navier--Stokes simulations. The viscosity law exponent $s=0.74$ was found to best match experimental data~\cite{Alsemeyer1976}. For further details about the DSMC simulations, readers are directed to~\cite{boyd_schwartzentruber_2017, valentini2009large}. We denote the target data  as $\overline{\rho}$, $\overline{u}$ and $\overline{T}$.

\subsection{Damped Newton Method} \label{sec: Newton solver}

Fully converging the forward solution is necessary to ensure well-posedness of the embedded optimization problem. This is achieved using a damped Newton method, which is able to converge the solution to near machine precision (using 64-bit floating-point precision).
By comparison,  pseudo-time-stepping methods require taking prohibitively many steps to achieve comparable convergence. For the present shock cases with moderately high Mach numbers (e.g., $M=5$) converging the relative loss to $O(10^{-2})$ requires approximately 1200 pseudo-time iterations but only approximately 20 damped Newton iterations. Even with the higher cost of the Newton solver, the lower number of iterations reduces the training time by approximately a factor of ten. 

Consider the steady solution of \eqref{eq:NS_1d}, $\partial(\bF_d-\bF_c)/\partial x = \partial\bF_x/\partial x=0$.
The $n^\mathrm{th}$ Newton iteration solves $\Delta\bU$ from the linearized system
\begin{equation}
    {\frac{d \bF_x}{d \bU}}\bigg|^{n} {\Delta \bU} = -\bF_x({\bU}^n) 
    \label{eq: Newt Iter}
\end{equation}
and advances the solution as $\bU^{n+1} = \bU^n + \Delta\bU$
until $\Delta \bU$ falls below a specified tolerance. Given the low dimensionality of the 1D problem, we solve \eqref{eq: Newt Iter} using Gauss--Jordan elimination, though 2D and 3D problems would require more sophisticated sparse or iterative solution methods. Since  the embedded optimization procedure currently uses the Jacobian ${{\partial \bF_x}/{\partial \bU}}$, the Jacobian can be reused for the Newton solver, though both could be posed in a Jacobian-free manner.

A practical difficulty with Newton iteration is the lack of physical constraints on $\Delta \bU$. Large values of $\Delta \bU$ obtained by solving \eqref{eq: Newt Iter} can result in nonphysical values of $\bU$ (e.g., negative density or pressure). We avoid these problems by bracketing $\Delta \bU$ such that it does not exceed a specified fraction of the local solution. The bracketing operation is
\begin{equation}
    {\Delta \bU_{b}} = \min\left( f_{b} \|{\bU}_{\infty} \|_2 \times \frac{\Delta {\bU}}{\| \Delta {\bU} \|_2},\ \Delta {\bU} \right),
\end{equation}
where $f_{b}=0.1$ is the bracketing factor, $\|{\bU}_{\infty} \|_2$ is the $L_2$ norm of the freestream $\bU$, and ${\Delta {\bU}}/{\| \Delta {\bU} \|_2}$ is the unit vector in the direction of $\Delta \bU$. To further ensure physical solutions, $ {\Delta \bU_{b}}$ is recursively damped, ${\Delta \bU_{d}} = f_{d} \Delta \bU_{b}$, where $ f_{d}=0.2$ is a damping factor,  until ${\bU}^{n+1}$ is physically consistent (e.g., the density and pressure are positive). The solution is then advanced as
\begin{equation}
    {\bU}^{n+1} = {\bU}^n + {\Delta \bU_{d}^n}.
    \label{eq: Newt Step damped}
\end{equation}
 Using the Newton method along with bracketing and damping stabilizes the solution before taking an optimization step to update the neural network parameters.

\section{Machine Learning Closure} \label{sec:ML}
This section is organized as follows: \ref{sec: Model architecture} outlines the architecture of the embedded neural network and its inputs, \ref{sec: Online Optimization} details the standard \emph{a priori} optimization approach and the adjoint-based optimization approach used in this paper, and \ref{sec:Ent_constr} presents a brief derivation of the entropy constraints introduced to enforce the second law of thermodynamics.

\subsection{Model Architecture} \label{sec: Model architecture}
The inputs to the embedded neural network are the first-order spatial derivatives of the computed scalar flow quantities and velocity at a mesh point $x_i$: $\hat z_i=\{\rho_x(x_i), p_x(x_i), T_x(x_i), u_x(x_i)\}$.
These inputs are invariant to Galilean transformations (constant relative motion), being spatial derivatives of scalar and vector quantities, which also guarantees that the network output quantities are invariant to Galilean transformations~\cite{wang2022extensions, ling2016machine}. We do not consider rotational invariance, which is a separate issue that has been assessed elsewhere \cite{ling2016machine}.
The model inputs are expanded to include data at the two neighboring mesh points: $z_i = \{\hat z_{i-1},\hat z_i,\hat z_{i+1}\}$. This expanded set of inputs serves to enable the model to compute higher derivatives, though it introduces a degree of mesh dependence, which could reduce generalizability to out-of-sample meshes. Addressing the issues of rotational invariance and mesh dependence is the subject of ongoing applications to two- and three-dimensional flows.

The neural network has a four-layer structure with one gate layer. For inputs $z\in\mathbb{R}^{18\times n_x}$, the model architecture is
\begin{equation*}
\begin{split}
  H^1 &= \sigma_1(W^1 z + b^1)   \\
  H^2 &= \sigma_1(W^2 H^1 + b^2)    \\
  H^3 &=  G \odot H^2 \qquad\qquad \text{with} \quad  G = \sigma_2(W^3 z + b^3) \\
  \hat f(z;{\theta}) &= W^4 H^3  + b^4
\end{split}
\end{equation*}
with output dimensions $\hat f(z;\theta)\in\mathbb{R}^{2\times n_x}$. In these,
\begin{equation*}
  \sigma_1(x) = \left\{\begin{array}{cc} \alpha(\exp(x)-1), & x\leq 0 \\ x, & x>0  \end{array}\right.
\end{equation*}
is the exponential linear unit (ELU) activation function with hyperparameter $\alpha=1$, $\sigma_2(x) = 1/(1 + e^{-x})$ is the sigmoid activation function, $\odot$ signifies element-wise multiplication, and the parameters are $\theta = \{ W^1, W^2, W^3, W^4, b^1, b^2, b^3, b^4 \}$. We use $N_H=1200$ hidden units per layer, which  balances training accuracy and computational efficiency.

For Approach B (augmented continuum viscous-stress and heat-flux models; \eqref{eq:aug_model}), the output layer is chosen to enforce the entropy conditions introduced in Section~\ref{sec:Ent_constr}. For Approach A (corrected transport coefficients; \eqref{eq:mu_aug_model}), the raw neural network outputs $\hat f(z;\theta)$ are passed through a final ELU layer,
\begin{equation*}
  f(z;\theta) = \sigma_1(\hat f(z;\theta)) + 0.1,
\end{equation*}
which is a simpler constraint to ensure positivity of $\mu_0(1 + f_1(z;\theta))$ and $\lambda_0(1 + f_2(z;\theta))$ for all $z$ and $\theta$.

\subsection{Online Optimization Approach} \label{sec: Online Optimization}

Optimization of standard \emph{a priori} DL closures is constrained to use objective (loss) functions that are explicit functions of the neural network. For example, if one wished to model the viscosity $\mu$ with a neural network $f(z;\theta)$, then it would be sufficient to minimize the objective function
\begin{equation}
  J(\theta) = \frac{1}{2}\int_{-\infty}^\infty \Big( f(z(x);\theta) - \mu^e \Big)^2\,dx,
  \label{eq:loss_apriori}
\end{equation}
where $\mu^e$ is trusted data (usually obtained from DNS), and $z(x)$ are the input variables. To select $\theta$ using gradient-descent optimization, one needs to compute the gradient
\begin{equation*}
  \nabla_\theta J = \int_{-\infty}^\infty \Big( f(z;\theta) - \mu^e \Big) \nabla_{\theta} f(z(x); \theta)\, dx,
\end{equation*}
where $\nabla_{\theta} f(z(x); \theta)$ may be evaluated using AD over the neural network, which is standard functionality in DL software libraries. However, optimization using \eqref{eq:loss_apriori} is problematic when data for the exact closure term ($\mu^e$) are not available---for example, when using experimental or higher-order simulation data (e.g., DSMC)---or when the target quantity is not known or accurately estimable from the available data. Furthermore, this optimization takes place offline, without solving the governing PDEs, so at best is only indirectly PDE-constrained by the target data.

The embedded DL approach addresses these shortcomings by introducing an online optimization algorithm~\cite{sirignano2021pdeconstrained,SirignanoOnlineAdjoint}, which enables arbitrarily defined objective functions. For the application to nonequilibrium flows, we construct an objective function using the mismatch between the computed primitive variables $\bu_p=[\rho,u,T]^\top$  and target data obtained from DSMC solutions of the distribution function.  Using an objective function constructed of primitive variables $\bu_p$ rather that conserved variables $\bU$ resulted in better convergence and lower losses. In general, the target data may be known from DNS, experiments, and/or higher-order simulations. The loss function in terms of the primitive variables is
\begin{equation} 
  J(\bu_p(\theta)) = \frac{1}{2} \int_{-\infty}^{\infty} \left[ \frac{1}{\rho_\infty^2}\left(\rho(t,x;\theta) - \overline{\rho}\right)^2 + \frac{1}{u_\infty^2}\left( u(t,x;\theta) - \overline{u}\right)^2 + \frac{1}{T_\infty^2}\left(T(t,x;\theta) - \overline{T}\right)^2 \right]\,dx,
  \label{eq:loss}
\end{equation}
where $\rho_\infty$, $u_\infty$, and $T_\infty$ are the reference freestream values used for normalization, and where we emphasize the dependence of the flow quantities on $\theta$.  The objective function depends on $\theta$ implicitly through $\bu_p$.

\begin{figure}
  \centering
  \includegraphics[width=0.75\textwidth]{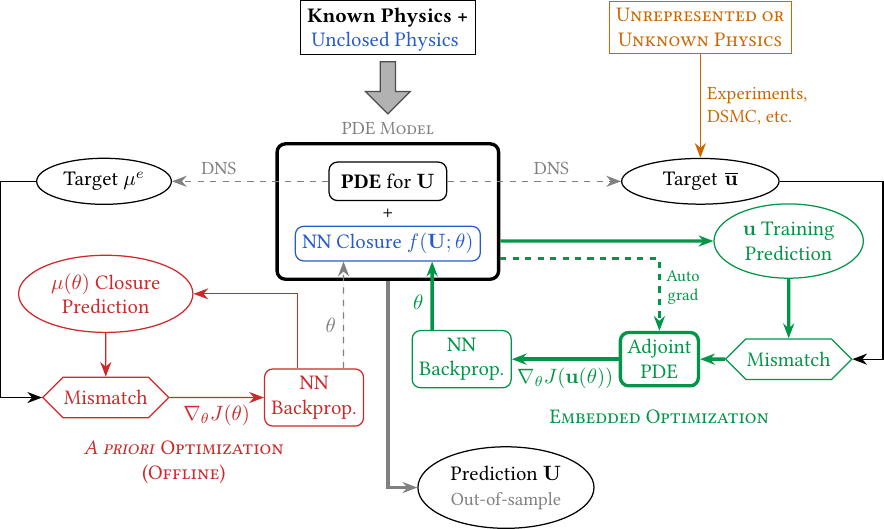}
  \caption{Two methods to optimize a NN closure $f(\bu;\theta)$ in a PDE model to represent unclosed physics. \emph{A priori} optimization selects the model parameters $\theta$ offline (without solving the PDE model). Embedded optimization instead minimizes the error with respect to the PDE solution $\bU$ (or derived quantities $\bu$) by solving adjoint PDEs, which enables efficient computation of the gradient $\nabla_\theta J(\bu(\theta))$. }
  \label{fig:DPM}
\end{figure}

Figure~\ref{fig:DPM} illustrates the conceptual differences between \emph{a priori} and embedded optimization. Notably, \emph{a priori} optimization requires high-fidelity (e.g., DNS) target data for the closure term, which may not be available or accurately known. (This would be the case, for example, for DNS solutions of the Navier--Stokes equations in the transition-continuum regime.) 
\emph{A priori} optimization cannot target the flow variables (or derived quantities), thus it cannot straightforwardly incorporate experimental and/or higher-order simulation data. In the same vein, the embedded optimization approach does not necessarily need DNS data.

We now consider the optimization of the discretized versions of (\ref{eq:NS_1d}) and (\ref{eq:loss}); all variables in the remainder of this section will therefore be finite-dimensional. This optimization problem is challenging, for it requires computing the gradients of dependent variables and derived quantities  with respect to the neural network parameters $\theta$. A na\"ive approach would attempt to evaluate
\begin{equation}
  \nabla_\theta J = \pp{\rho}{\theta}^{\top} \pp{J}{\rho} + \pp{u}{\theta}^{\top} \pp{J}{u} + \pp{T}{\theta}^{\top} \pp{J}{T} = \pp{\bu_p}{\theta}^{\top} \pp{J}{\bu_p}
  \label{eq:grad_J}
\end{equation}
on the discrete computational mesh. The partial derivatives $\partial\bu_p/\partial\theta\in\mathbb{R}^{3n_x\times n_\theta}$ could be evaluated by deriving and solving PDEs for each $\partial\bu/\partial\theta_i$, where $i = 1, \ldots, n_{\theta}$, or by introducing finite-difference perturbations $\theta_i+\theta'_i$. However, the number of neural network parameters $n_\theta$ is often large (e.g., $\sim 10^5$ parameters), such that both of these approaches would require an intractable number of calculations. 

Instead, for $n$ forward equations, $n$ adjoint equations may be solved to provide the necessary gradients. Thus the overall cost will not be significantly higher for optimization than for prediction. To derive the adjoint equations for a steady problem, note that
\begin{equation}
  \nabla_\theta\mathbf{F}_x = \pp{\mathbf{F}_x}{\bu_p}\pp{\bu_p}{\theta} + \pp{\mathbf{F}_x}{\theta} = 0,
  \label{eq:grad_F}
\end{equation}
where $\mathbf{F}_x$ is the discrete representation of the right-hand side (RHS) $\frac{\partial}{\partial x}(\mathbf{F}_d - \mathbf{F}_c)$.

Let us now introduce the adjoint variables $\hat\bu = [\hat \rho, \hat u, \hat T]^\top$, which satisfy the adjoint equation 
\begin{equation}
  \pp{J}{\bu_p} + \pp{\bF_x}{\bu_p}^{\top} \hat\bu = 0 ,
  \label{eq:adj1}
\end{equation}
with Dirichlet boundary conditions $(\hat \rho, \hat u, \hat T) = 0$ at $x = \pm \infty$. 
Multiplying (\ref{eq:grad_F}) by $\hat \bu^{\top}$ produces
\begin{equation}
0 = \hat \bu^{\top} \pp{\mathbf{F}_x}{\bu_p}\pp{\bu_p}{\theta} + \hat \bu^{\top} \pp{\mathbf{F}_x}{\theta}.
\label{eq:ForwardMultiplied}
\end{equation}
From (\ref{eq:adj1}), we have that $\hat\bu^{\top} \pp{\bF_x}{\bu_p}  = - \pp{J}{\bu_p}^{\top} $ which, when substituted into (\ref{eq:ForwardMultiplied}), yields
\begin{equation}
\pp{J}{\bu_p}^{\top} \pp{\bu_p}{\theta} = \hat \bu^{\top} \pp{\mathbf{F}_x}{\theta}.
\label{eq:JMultiplied}
\end{equation}
The relationship \eqref{eq:JMultiplied} combined with (\ref{eq:grad_J}) provides a formula for the gradient in terms of the adjoint solution:
\begin{equation}
\nabla_\theta J  = \pp{\mathbf{F}_x}{\theta}^{\top}  \hat \bu. 
\label{eq:grad_J_adj}
\end{equation}
Therefore, we can solve the adjoint equation (\ref{eq:adj1}) for $\hat\bu$ and use its solution to evaluate the gradient of the loss function with respect to the model parameters ($\nabla_\theta J$) using \eqref{eq:grad_J_adj}. The dimension of \eqref{eq:adj1} is equal to the dimension of the forward equations \eqref{eq:NS_1d}, so the overall optimization is not significantly greater than the cost of solving the forward equations. It is important to note that the validity of \eqref{eq:grad_J_adj} is contingent on the right-hand side of \eqref{eq:grad_F} being zero. In turn, this requires that the discrete residual of the forward equations is close to zero before taking an optimization step. 

The discrete adjoint equations are solved as a linear system. This requires the computation of the Jacobian ${\partial\bF_x}/{\partial\bu_p}$, which is calculated using  algorithmic differentiation as provided by the \emph{PyTorch} library~\cite{PyTorchNIPS}. After converging the forward and adjoint equations, gradient-descent steps are taken for parameters $\theta^{(k)}$ according to 
\begin{equation}
  \theta^{(k+1)} = \theta^{(k)} - \alpha^{(k)}  \nabla_\theta J(\theta^{(k)}),
  \label{eq:param_update}
\end{equation}
where $\alpha^{(k)}$ is the learning rate at the iteration $k\in \{0, 1, \ldots, N_\mathrm{iter} \}$, where $N_\mathrm{iter}$ is the maximum number of optimization iterations. The learning-rate schedule and other model-training hyperparameters are described in Section \ref{sec:ML_train}.

\subsection{Entropy Constraints} \label{sec:Ent_constr}

Using DL closures unconstrained for the second law of thermodynamics can result in unphysical flow quantities. We constrain the outputs of the neural networks to satisfy the second law of thermodynamics using the Clausius--Duhem inequality~\cite{powers2016combustion},
\begin{equation} 
    \mathcal{\Dot{I}} = -\frac{1}{T^2} \bq(\bU;\theta) \nabla T + \frac{1}{T} \sigma(\bU;\theta) : \nabla \bu^T \ge 0,
    \label{eq:irrRevEnt}
\end{equation}
where $\mathcal{\Dot{I}}$ is the rate of irreversible entropy production. A ``strong'' form of the entropy constraint would impose conditions on the model outputs such that (\ref{eq:irrRevEnt}) is nonnegative. For Approach A (transport-coefficient corrections), ensuring the positivity of $\mu(\bU;\theta)$ and $\lambda(\bU;\theta)$ is sufficient; this is done using an ELU output layer as described in Section~\ref{sec:ML}. A ``weak'' constraint for the entropy inequality would incorporate $\mathcal{\Dot{I}}$ as a penalty term in the training loss function \eqref{eq:loss}.  The application of strong and weak constraints for Approach B (transport-term augmentation) is now discussed separately.

\subsubsection{Strong Entropy Constraint}

The Clausius--Duhem inequality \eqref{eq:irrRevEnt} can be rearranged to give
\begin{equation}
    \frac{1}{T} \sigma(\bU;\theta) : \nabla \bu^T \ge -\frac{1}{T^2} \bq(\bU;\theta) \nabla T,
\end{equation}
from which a constraint on the viscous stress may be obtained:
\begin{equation}
    \sigma(\bU;\theta) \ge \frac{\frac{1}{T} \bq(\bU;\theta)  \nabla T}{ \nabla \bu^T}.
\end{equation}
A constrained viscous-stress model is therefore
\begin{equation}
    \tilde{\sigma}(\bU;\theta) = \max(\sigma(\bU;\theta), \mathbf{N}),
    \label{eq: Strong Ent sigma}
\end{equation}
where 
\begin{equation}
   \mathbf{N} =  \frac{\frac{1}{T} \bq(\bU;\theta)  \nabla T}{ \nabla \bu^T},
    \label{eq: N strong ent}
\end{equation}
and the max() function in \eqref{eq: Strong Ent sigma} is applied in a pointwise manner. After applying \eqref{eq: Strong Ent sigma}, the constrained closure terms $\bq(\theta)$ and $\tilde\sigma(\theta)$ are substituted into the governing equations. This enforcement of the strong entropy condition makes no assumptions about the signs of $\nabla T$ and $\nabla u$.

For 1D flows, the Approach B closures \eqref{eq:aug_model} can be simplified to
\begin{align}
  \begin{split}
    \sigma(\bU;\theta) &= \frac{4}{3} \mu \frac{\partial u}{\partial x} - f_1(\textbf{U}_x;\theta)\qquad \\
    \bq(\bU;\theta) &= -\lambda \frac{\partial T}{\partial x} - f_2(\textbf{U}_x;\theta),
  \end{split}
\end{align}
and the Clausius-Duhem inequality \eqref{eq:irrRevEnt} reduces to
\begin{equation} 
  \mathcal{\Dot{I}} = -\frac{1}{T^2} \bq(\bU;\theta) \frac{\partial T}{\partial x} + \frac{1}{T} \sigma(\bU;\theta) \frac{\partial u}{\partial x} \ge 0.
  \label{eq: Clausis-duhem 1D}
\end{equation}
The term $\mathbf{N}$ \eqref{eq: N strong ent} reduces to
\begin{equation}
  \mathbf{N} =  \frac{\frac{1}{T} \bq(\bU;\theta)  \pp{T}{x}}{ \pp{u}{x}},
\end{equation}
and the strong entropy constraint can be enforced according to \eqref{eq: Strong Ent sigma}.

\subsubsection{Weak Entropy Constraint}
A weak entropy constraint would penalize the violation of the entropy inequality \eqref{eq:irrRevEnt} during the constrained optimization process by adding an extra term to the training loss function \eqref{eq:loss},
\begin{equation}
    J_{s}(\theta) =  W_{s}\int_{-\infty}^{\infty}\left[ \min(0,f_1(\textbf{U}_x;\theta)) +  \min(0,f_2(\textbf{U}_x;\theta)) \right] d \bx,
    \label{eq: Weak ent loss}
\end{equation}
where $W_{s}$ is a scalar weight. The net training loss for the weak entropy constraint is thus
\begin{equation}
    J_w(\bu_p;\theta) = J(\bu_p(\theta)) + J_{s}(\bu_p(\theta)).
\end{equation}
The requirement to simultaneously minimize both components of $J_w$ prevented weak-constraint cases from matching the testing and training accuracy of equivalent strong-constraint cases. This could be improved by dynamic selection of $W_s$, for example, to enforce the $J_s$ term more strongly in regions of high entropy production.

\subsection{Model Training} \label{sec:ML_train}
A distributed training approach was implemented to perform the online adjoint optimization. 
For each of Approaches A and B, models were trained for four different training regimens: three models targeted only a single freestream Mach number ($M_\mathrm{train}=2$, $M_\mathrm{train}=5$, and $M_\mathrm{train}=8$), and one model targeted three Mach numbers simultaneously: $M_\mathrm{train}=(2,5,8)$. Training this combined model utilized message-passing interface (MPI) parallelization to average the loss-function gradients $\nabla_\theta J$ between the component simulations before optimizing.

Table~\ref{tab:train} provides the training/testing regimen matrix. Freestream Mach numbers $M_\infty=2,5,8$ were out-of-sample when not included in a model's training regimen. Freestream Mach numbers $M_\infty=3,6,9,10$ were out-of-sample for all trained models. This testing arrangement is designed to evaluate the stability and accuracy of the trained models for interpolation ($M_\mathrm{train}=(2,5,8)$ only) and extrapolation to lower and higher Mach numbers.
\begin{table}
  \centering
  \begin{tabular}{c c c c c c c c}
    \toprule
    Training & $M_\infty=2$ & $M_\infty=3$ & $M_\infty=5$ & $M_\infty=6$ & $M_\infty=8$ & $M_\infty=9$ & $M_\infty=10$ \\
    \midrule
    $M_\mathrm{train}=2$ & $\bullet$ \\
    $M_\mathrm{train}=5$ & & & $\bullet$ \\
    $M_\mathrm{train}=8$ & & & & & $\bullet$ \\
    $M_\mathrm{train}=(2,5,8)$ & $\bullet$ & & $\bullet$ & & $\bullet$ \\
    \midrule
    Testing & $\circ$ & $\bullet$ & $\circ$ &  $\bullet$ & $\circ$ &  $\bullet$ &  $\bullet$ \\
    \bottomrule
  \end{tabular}
  \caption{Model training and testing matrix. $M_\infty=2,5,8$ ($\circ$) were out-of-sample when not used for training. $M_\infty=3,6,9,10$ ($\bullet$) were out-of-sample for all models.}
  \label{tab:train}
\end{table}

Initial Navier--Stokes viscous-shock solutions for each freestream Mach number listed in Table~\ref{tab:train} were obtained by converging from the corresponding inviscid-shock profile. Training and testing cases were initialized from the converged Navier--Stokes solutions. For model training, the parameters of randomly-initialized neural networks were selected using \eqref{eq:param_update} with initial learning rate $\alpha^{(0)}=2.5\times 10^{-6}$. To ensure convergence, the learning rate was reduced using an adaptive schedule:
\begin{equation*}
  \alpha = \left\{\begin{array}{cc} \alpha, & \relErr > \epsilon_T \\ 0.75\alpha, & \relErr\leq\epsilon_T \end{array}\right., \qquad
  \epsilon_T = \left\{\begin{array}{cc} \epsilon_T, & \relErr > \epsilon_T \\ 0.75\epsilon_T, & \relErr\leq\epsilon_T \end{array}\right.,
\end{equation*}
where $\relErr = J(\theta)/J_0$ is the objective-function relative error, $J_0$ is the initial value of the objective function (i.e., for the initial, unmodified Navier--Stokes solution), and the starting target error is $\epsilon_T=0.9$. For models trained simultaneously for $M_\mathrm{train}=(2,5,8)$, the minimum  $\alpha$ and  $\epsilon_T$ were synchronized between training cases. The training process was stopped for change in relative error $\relErr$ less than $1 \times 10^{-5}$ between successive iterations.  Algorithm \ref{alg:Training} summarizes the training process used for all training regimens.

\begin{algorithm}
\caption{Training process}\label{alg:Training}
\begin{algorithmic}[1]

\State Restart $\bU$ from unmodified Navier--Stokes solution
\State Randomly initialize DL network parameters $\theta^0$
\State Calculate loss of unmodified solution $J_0$
\State Set $\epsilon_\mathrm{rel}^0 = 1.0$
\While{$\Delta \epsilon_\mathrm{rel}^{n} \ge 10^{-5}$}
    \State Converge solution to steady-state using damped Newton solver \ref{sec: Newton solver}
    \State Solve adjoint equation \eqref{eq:adj1} for $\hat{\mathbf{u}}$
    \State Compute  $\nabla_{\theta} J(\bu_p(\theta^n))$ using \eqref{eq:grad_J_adj}
    \State Update parameters $\theta$ using gradient descent \eqref{eq:param_update}
    \State Calculate $\epsilon_\mathrm{rel}^{n+1}$
    \State Compute $\Delta \epsilon_\mathrm{rel}^{n+1} = \epsilon_\mathrm{rel}^{n+1} - \epsilon_\mathrm{rel}^{n}$
\EndWhile
\end{algorithmic}
\end{algorithm}

Table~\ref{tab:loss} shows the total number of optimization iterations and  the average number of Newton iterations per optimization iteration for all models.
For all training cases, the relative tolerance of the forward Newton solver  was $1 \times 10^{-15}$ before proceeding to optimization. For both Approach A and Approach B, the higher Mach number cases required more Newton iterations per optimization iteration. Approach A resulted in better training accuracy (lower $\relErr$) for most training cases.
\begin{table}
  \centering
  \begin{tabular}{c c c c}
    \toprule
    Training & $N_\mathrm{iter}^\mathrm{opt}$ & $N_\mathrm{iter}^\mathrm{Newt}$ & $\epsilon_\mathrm{rel}$  \\
    \midrule
    \textit{Approach A} \\
    $M_\mathrm{train}=2$ & 21 & 8 & $5.333 \times 10^{-2}$  \\
    $M_\mathrm{train}=5$ & 19 & 20 & $1.774 \times 10^{-2}$ \\
    $M_\mathrm{train}=8$ & 21 & 35 & $1.069 \times 10^{-2}$ \\
    $M_\mathrm{train}=(2,5,8)$ & (100, 100, 100) & (9, 36, 17) & ($6.715 \times 10^{-1}$, $9.487 \times 10^{-2}$, $6.099 \times 10^{-2}$)  \\
    \midrule
    \textit{Approach B} \\
    $M_\mathrm{train}=2$ & 46 & 9 & $2.876 \times 10^{-2}$  \\
    $M_\mathrm{train}=5$ & 37 & 19 & $4.855 \times 10^{-2}$ \\
    $M_\mathrm{train}=8$ & 22 & 41 & $7.334 \times 10^{-2}$ \\
    $M_\mathrm{train}=(2,5,8)$ & (100, 100, 100) & (6, 22, 33) & ($4.264 \times 10^{-1}$, $7.516 \times 10^{-2}$, $5.397 \times 10^{-2}$)  \\
    \bottomrule
  \end{tabular}
  \caption{Number of optimization iterations, $N_\mathrm{iter}^\mathrm{opt}$, average number of Newton iterations per optimization iteration $N_\mathrm{iter}^\mathrm{Newt}$, and training loss (objective-function relative error $\epsilon_\mathrm{rel}$) for Approaches A and B. For $M_\mathrm{train}=(2,5,8)$, the training loss is listed separately for the three target Mach numbers.}
  \label{tab:loss}
\end{table}

\section{Results and Analysis} \label{sec:results}

\subsection{Shock Predictions}
\begin{figure}
  \centering
  \includegraphics[width=0.49\textwidth]{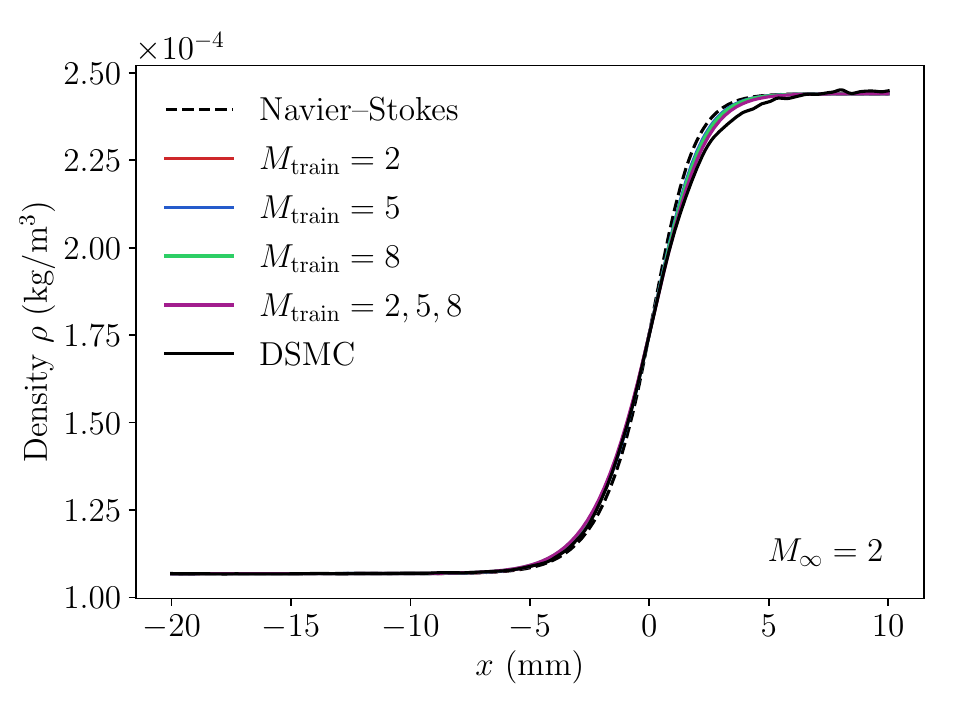}
  \includegraphics[width=0.49\textwidth]{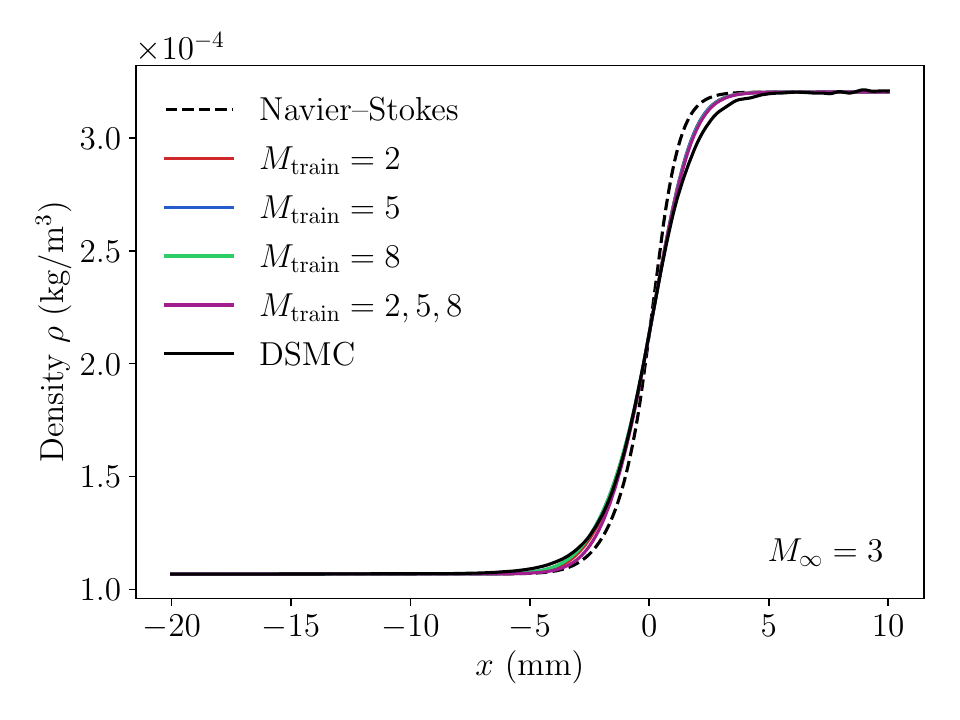}
  \includegraphics[width=0.49\textwidth]{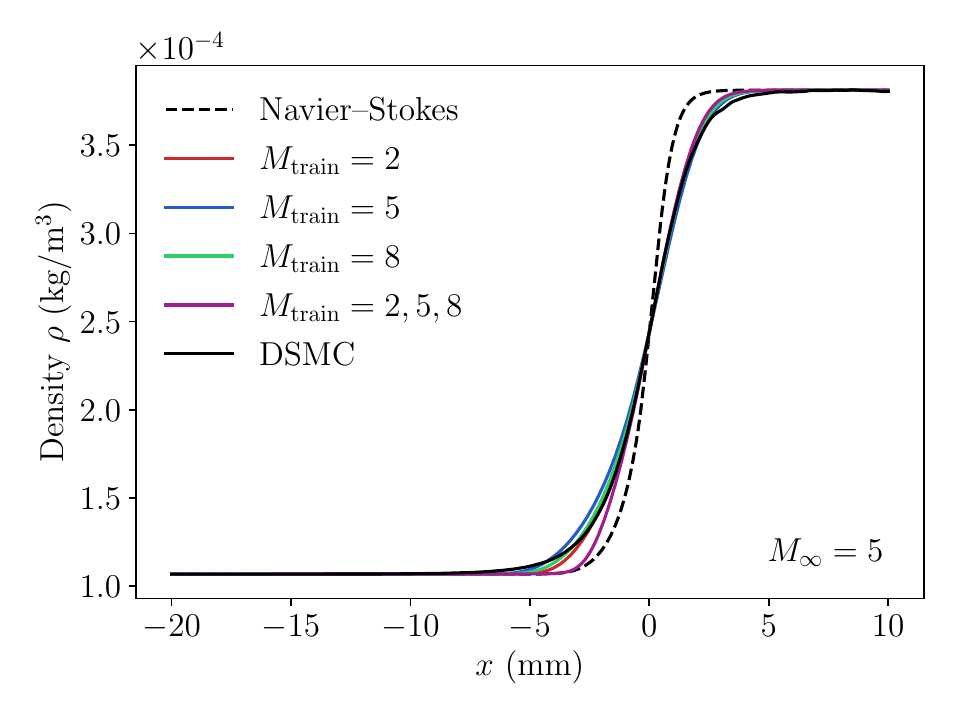}
  \includegraphics[width=0.49\textwidth]{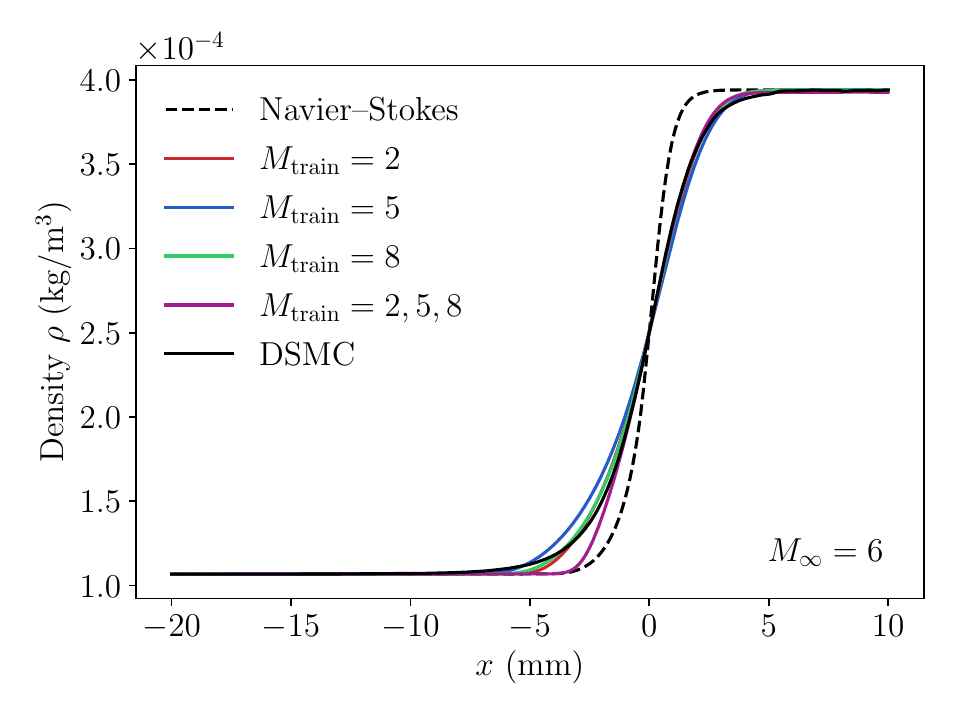}
  \includegraphics[width=0.49\textwidth]{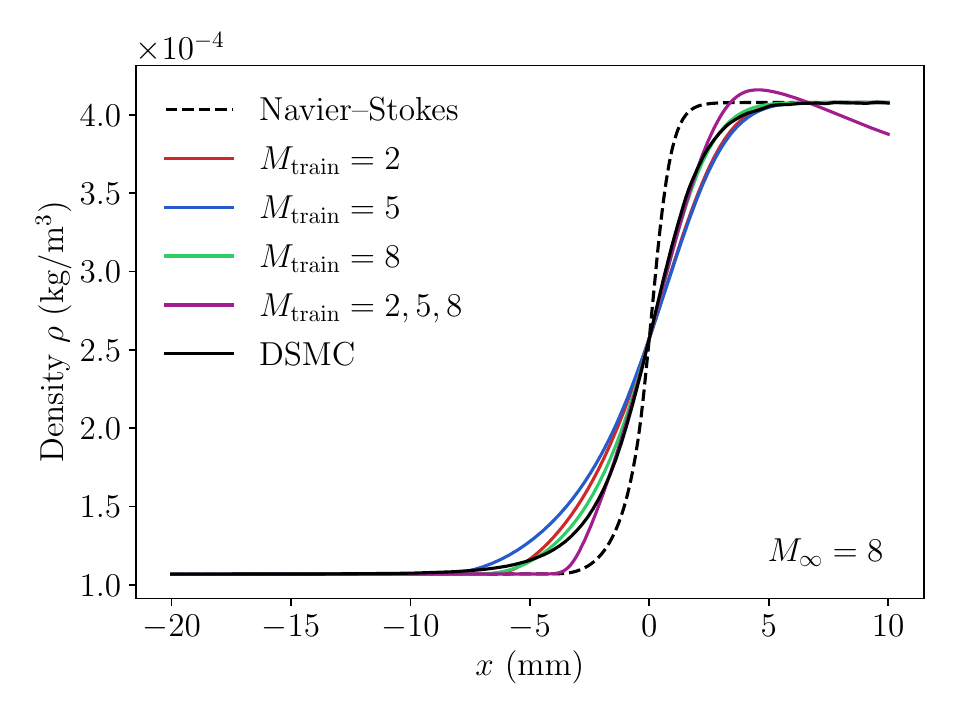}
  \includegraphics[width=0.49\textwidth]{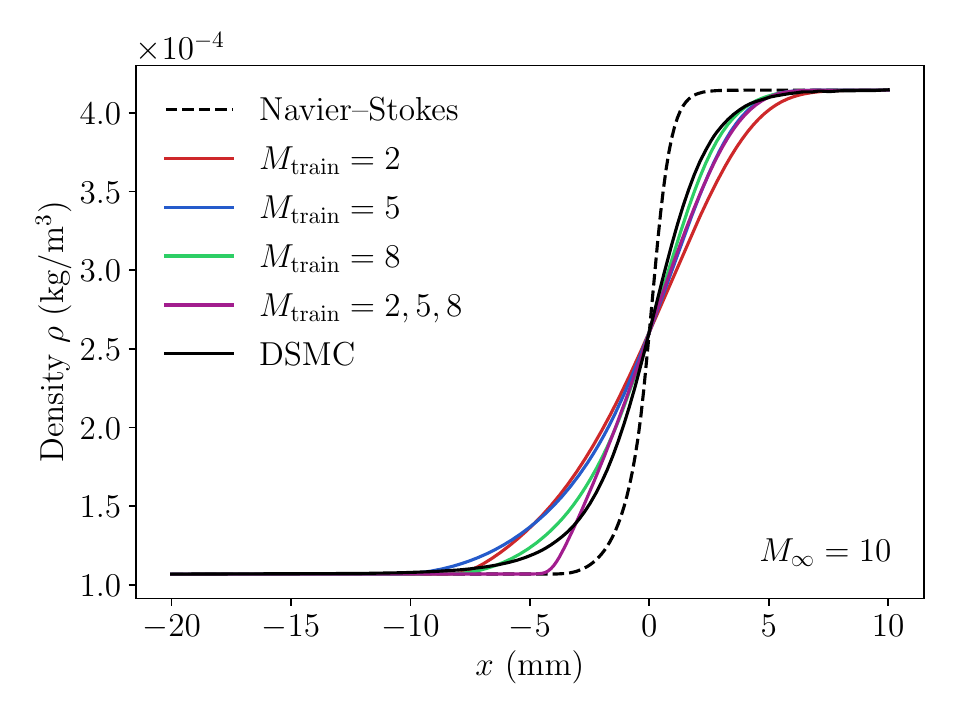}
  \caption{Approach A: In- and out-of-sample freestream density predictions. Results are shown for the unmodified Navier--Stokes equations, the Navier--Stokes equations with DL models for Approach A trained for $M=2$, $M=5$, $M=8$, and $M=(2,5,8)$, and the DSMC target data. The $M=3$, $M=6$, and $M=10$ cases are out-of-sample for all trained DL models.}
  \label{fig:rho_coeff}
\end{figure}

\begin{figure}
  \centering
  \includegraphics[width=0.49\textwidth]{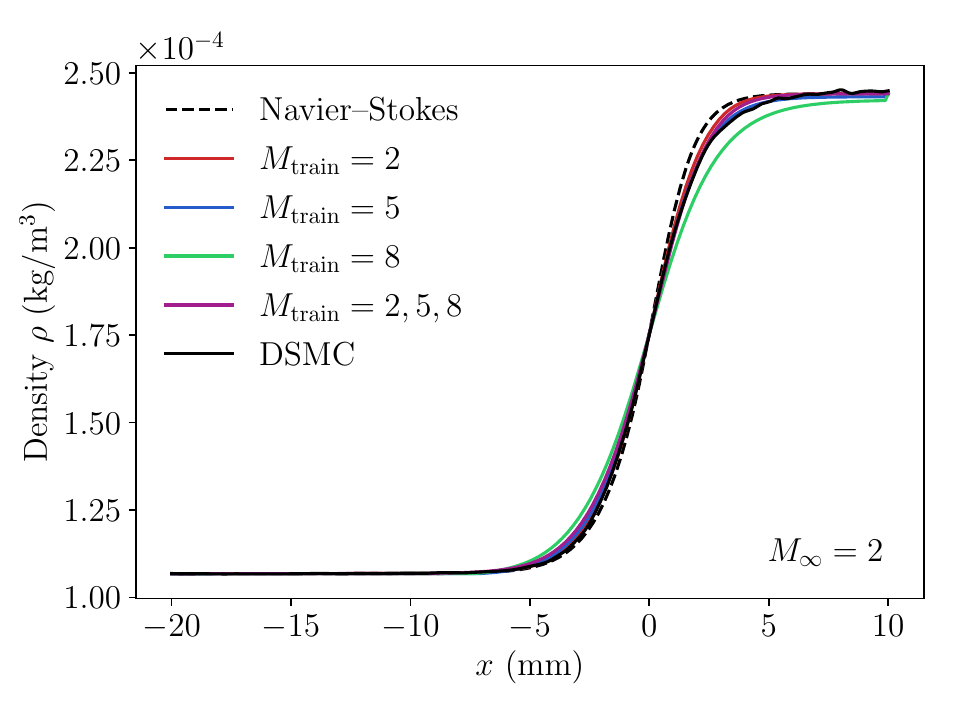}
  \includegraphics[width=0.49\textwidth]{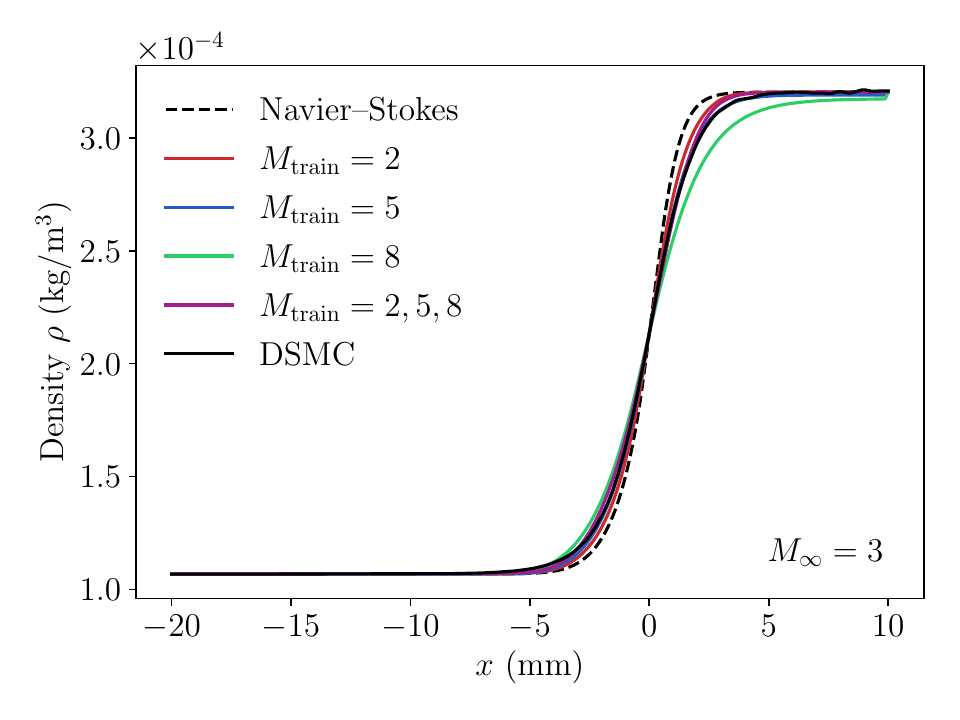}
  \includegraphics[width=0.49\textwidth]{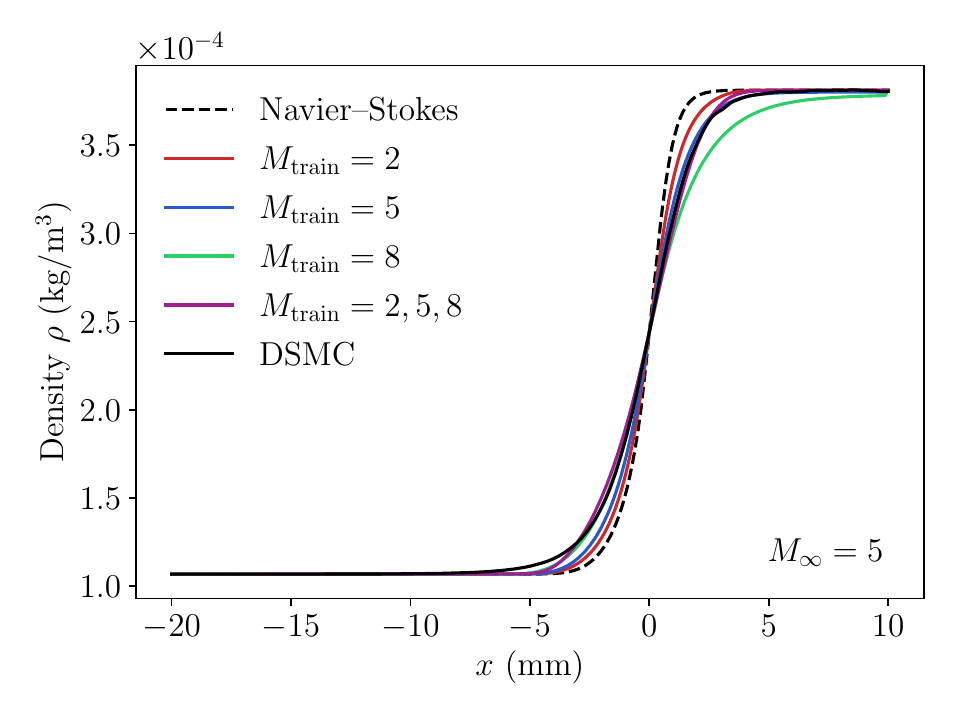}
  \includegraphics[width=0.49\textwidth]{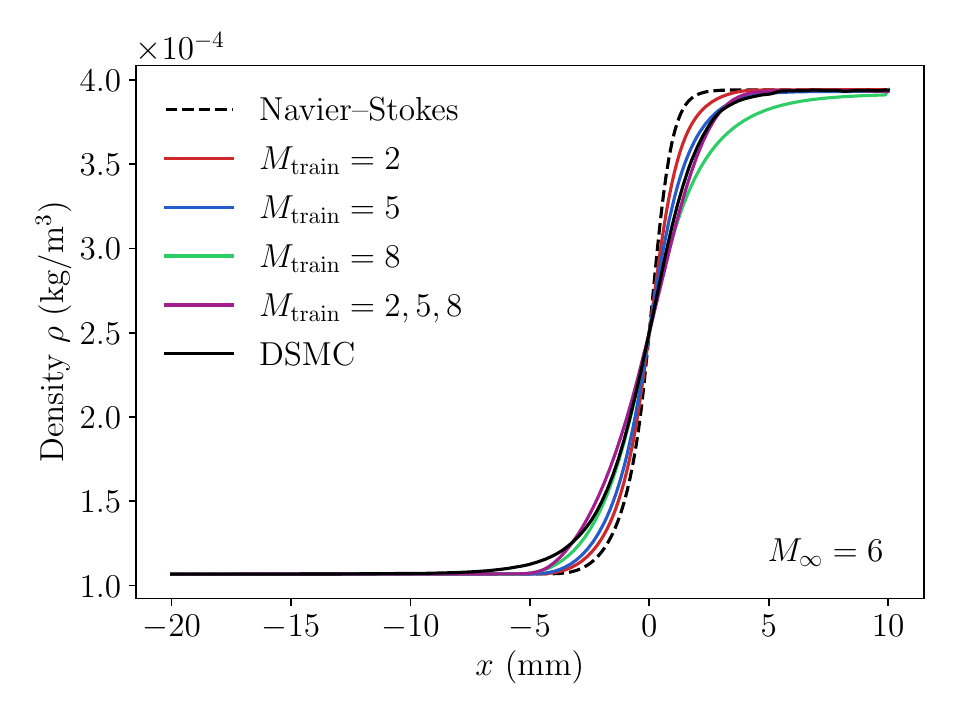}
  \includegraphics[width=0.49\textwidth]{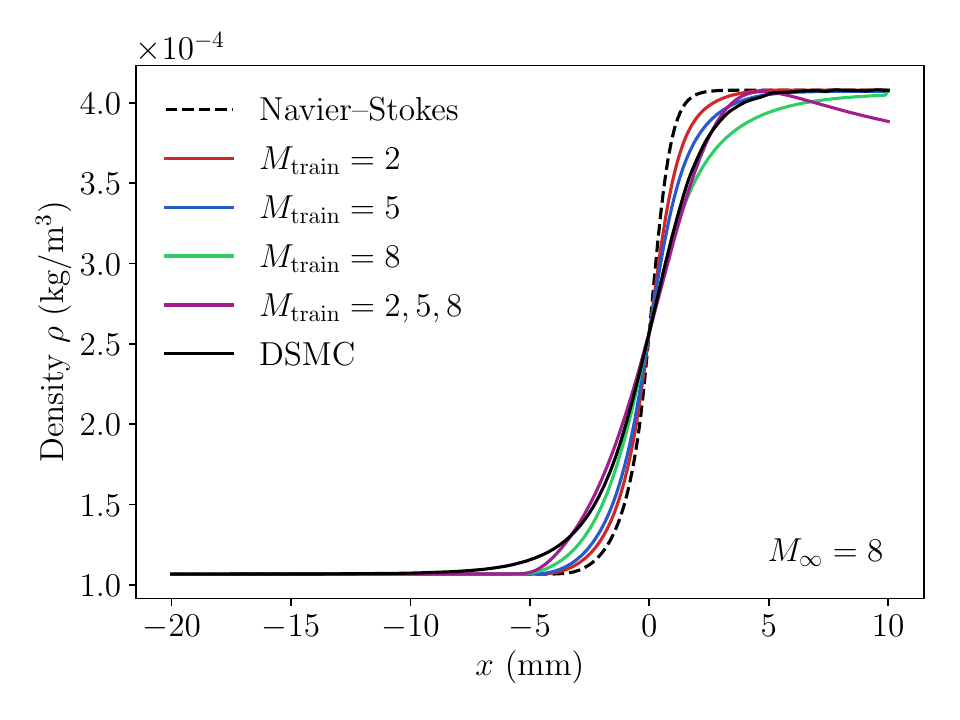}
  \includegraphics[width=0.49\textwidth]{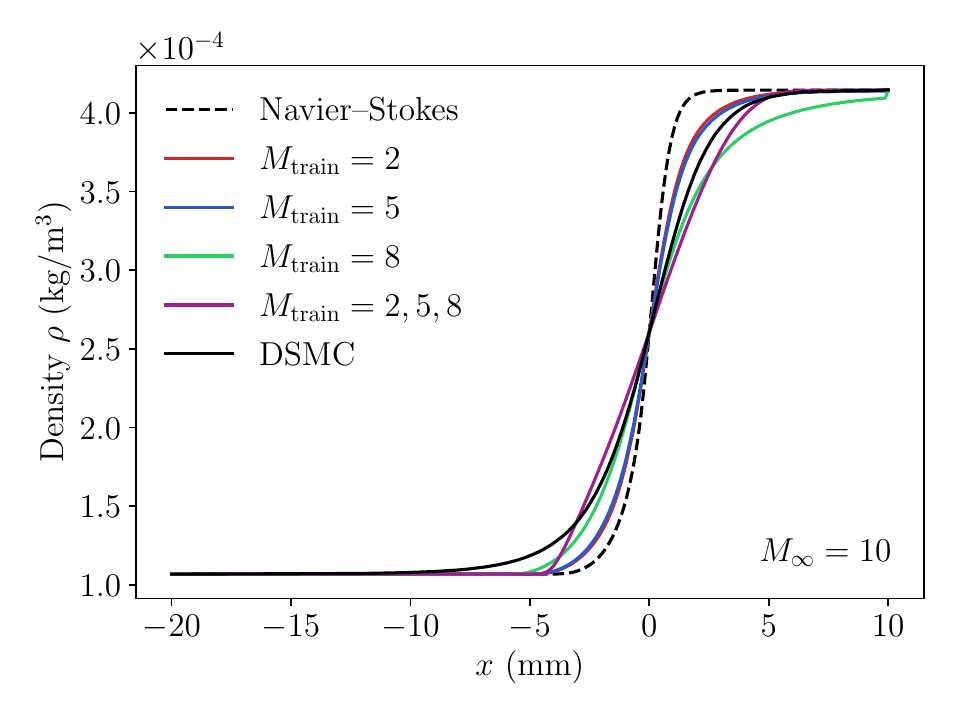}
  \caption{Approach B: In- and out-of-sample freestream density predictions. Results are shown for the unmodified Navier--Stokes equations, the Navier--Stokes equations with DL models for Approach B trained for $M=2$, $M=5$, $M=8$, and $M=(2,5,8)$, and the DSMC target data. The $M=3$, $M=6$, and $M=10$ cases are out-of-sample for all trained DL models.}
  \label{fig:rho_strong}
\end{figure}

Out-of-sample results were obtained for the testing cases listed in Table~\ref{tab:train}.  Without a closure model, the Navier--Stokes equations give qualitatively incorrect predictions for all training and testing cases.
Figures~\ref{fig:rho_coeff} and \ref{fig:rho_strong} (Approaches A and B, respectively) show the computed density from the unmodified Navier--Stokes equations, the Navier--Stokes solutions with the trained neural networks, and the target DSMC solutions for $M_\infty\in[2,10]$. 
The neural network-augmented Navier--Stokes solutions more closely match the target data for all cases and models. As could be anticipated from the training loss (Table~\ref{tab:loss}), the DL-augmented Navier--Stokes agreement with the DSMC data is qualitatively better for lower Mach numbers and for Mach numbers closer to each model's training conditions. It is also noteworthy that, for both approaches, the models trained on higher Mach numbers extrapolate better to lower Mach numbers. Similar trends were obtained for the pressure, temperature, and velocity fields.

\subsection{Prediction Accuracy} \label{sec:results_cvg}
\begin{figure}
  \centering
  \includegraphics[width=0.49\textwidth]{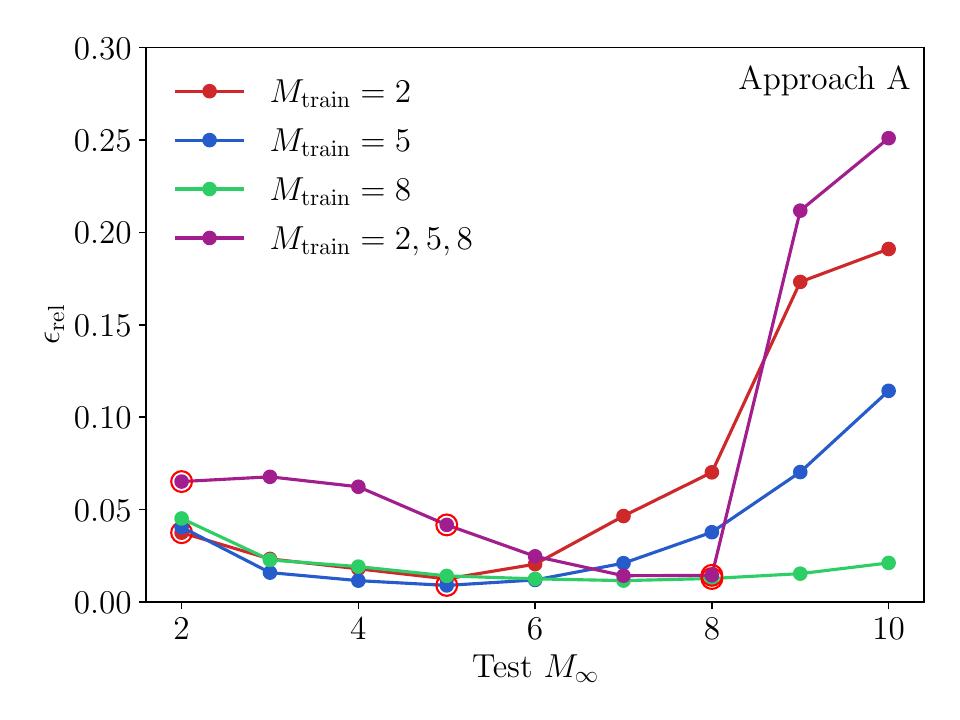}
  \includegraphics[width=0.49\textwidth]{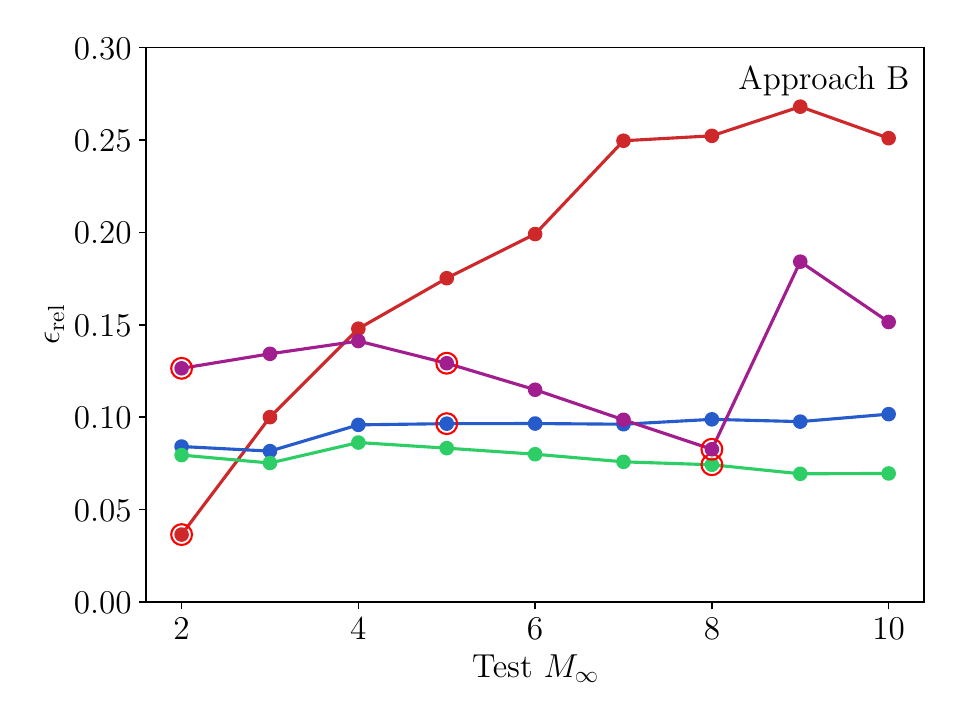}
  \caption{ In- and out-of-sample objective-function relative error for the four DL models trained for Approach A (left) and Approach B (right). In-sample cases for each model are indicated by red circles. The unmodified Navier--Stokes solution has $\relErr=1.0$.}
  \label{fig:loss_strong_coeff}
\end{figure}

Figure \ref{fig:loss_strong_coeff} shows the relative loss-function error as a function of the testing Mach number. This represents the integrated error of the primitive variables over the entire domain and can be interpreted as a worst-case assessment of a model's performance. All models trained for a single Mach number using Approach A perform better (lower $\relErr$) than the corresponding models using the Approach B. For  Approach B, the model trained for $\Mtr=2$ is generally more accurate at lower than higher testing Mach numbers, and the model trained for $\Mtr=8$ is generally more accurate at higher than lower testing Mach numbers, as could be anticipated. The Approach B model trained simultaneously for $\Mtr=(2,5,8)$ is reasonably accurate across the range of testing Mach numbers, with comparable accuracy to the in-sample Mach 2, 5, and 8 models, though its extrapolation accuracy diminishes at $M_\infty=9,10$. Interestingly, the single-case $\Mtr=8$ model reached the best in-sample local minimum and gave the lowest relative errors for lower testing Mach numbers. This model is also reasonably accurate for higher testing Mach numbers.

The extrapolation trends of Approach A are similar to those of Approach B. The model trained for $\Mtr=8$ extrapolates well to lower Mach number cases, with comparable accuracy to the $\Mtr=2$ case. However, the model trained simultaneously for $\Mtr=(2,5,8)$ performs poorly for this case with errors higher than all other models for almost all Mach numbers. The model trained simultaneously for for $\Mtr=(2,5,8)$ is reasonably accurate for $M_\infty=2$ through 8 but has increased errors when extrapolating to the higher $M_\infty=9$ and 10.

\subsection{Shock Thickness}

\begin{figure}
  \centering
  \includegraphics[width=0.49\textwidth]{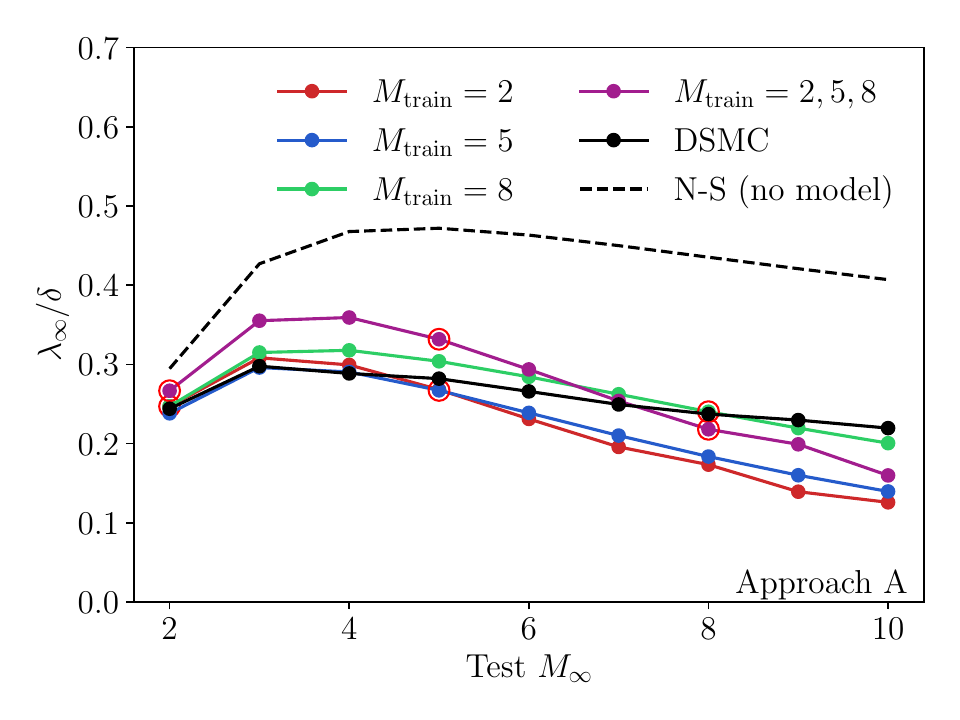}
  \includegraphics[width=0.49\textwidth]{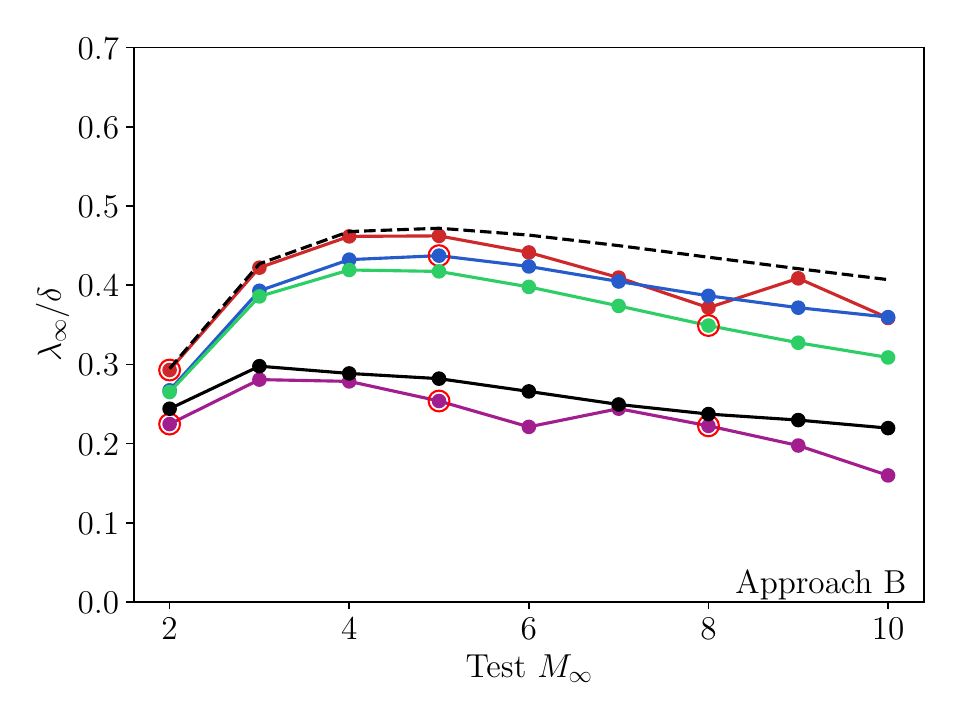}
  \caption{Predicted inverse shock thickness from the unmodified (no-model) Navier--Stokes equations, the augmented NS equations for the four DL models, and the DSMC target data for  Approach A (left) and Approach B (right).}
  \label{fig:thk_strong_coeff}
\end{figure}

The shock thickness is evaluated using the maximum density gradient,
\begin{equation}
  \delta = \frac{\max(\rho)-\min(\rho)}{\max(\partial \rho / \partial x)}.
\end{equation}
Figure~\ref{fig:thk_strong_coeff} plots the inverse shock thickness normalized by the mean free path~\cite{Alsemeyer1976},
\begin{equation}
  \lambda_{\infty} = \frac{16}{5} \left(\frac{\gamma}{2 \pi} \right)^{\frac{1}{2}} \frac{\mu_0}{\rho_{\infty} a_{\infty}},
\end{equation}
where $a_{\infty}$ is the freestream speed of sound and $\lambda_{\infty}=1.098$~mm for the temperature and pressure conditions considered here~\cite{Alsemeyer1976}.
As shown in Fig.~\ref{fig:thk_strong_coeff}, all of the DL models improve the Navier--Stokes predictions' accuracy over the unmodified continuum solutions, which overpredict the shock thickness for all freestream Mach numbers at this very low freestream density. The models trained for Approach A are generally more accurate than those trained for Approach B, which is counterintuitive, as one would expect the greater modeling flexibility of Approach B to result in higher overall accuracy, though this is generally consistent with the $\relErr$ trends (Fig.~\ref{fig:loss_strong_coeff}). The accuracy reduction for Approach B is likely due to inconsistencies in the Clausius--Duhem inequality when applied as a constraint for subcontinuum flows. Conversely, for Approach A, the simple positivity constraint on $\mu$ and $k$ is a stricter requirement but does not necessarily require the validity of the Chapman--Enskog expansion of the Boltzmann equation. For both Approaches A and B, the models trained simultaneously  for $\Mtr=(2,5,8)$ is reasonably accurate for shock-thickness predictions, especially at their in-sample conditions. The Approach A model trained for $\Mtr=8$ gives the most accurate inverse shock thickness predictions over the entire range of testing Mach numbers and is used for subsequent comparisons.

\begin{figure}
  \centering
  \includegraphics[width=0.6\textwidth]{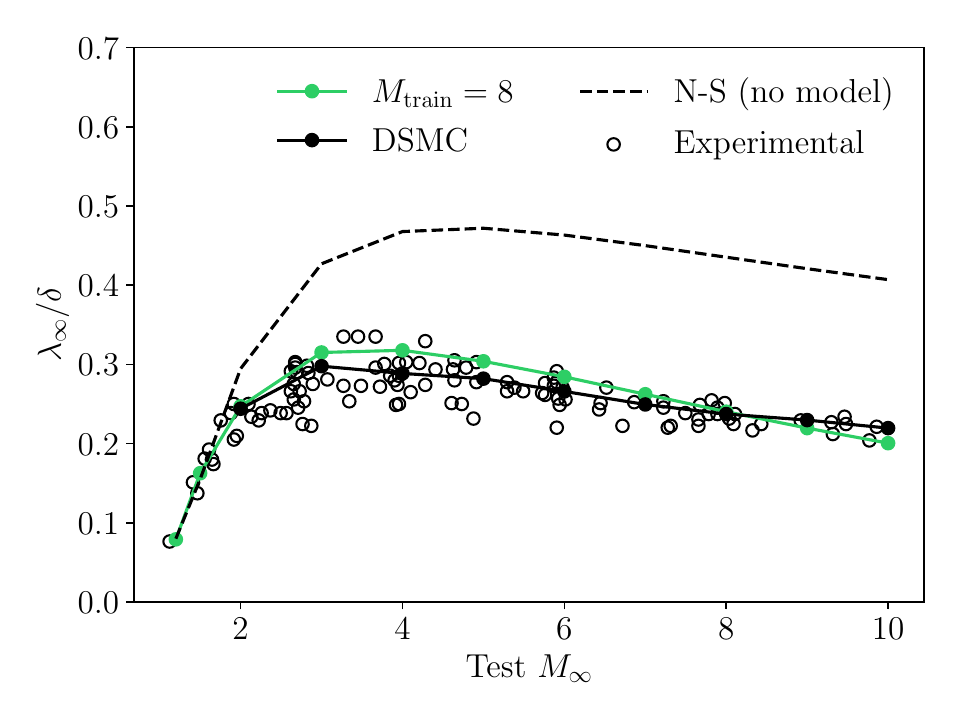}
  \caption{Experimental data from \cite{LiznerAndHornig, Camac, Schmidt1969, Garen1974, Alsemeyer1976} compared to predicted inverse shock thickness from the unmodified (no-model) Navier--Stokes equations, the augmented NS equations for $M_\mathrm{train}=(2,5,8)$ using Approach A, and the DSMC target data. }
  \label{fig:thk_strong_coeff_expt}
\end{figure}

Figure~\ref{fig:thk_strong_coeff_expt} compares the shock-thickness predictions of the Approach A, $\Mtr=8$ model to the DSMC targets and experimental data~\cite{LiznerAndHornig, Camac, Schmidt1969, Garen1974, Alsemeyer1976}. The DL model follows the DSMC trends to within the experimental uncertainty. Additionally, Fig.~\ref{fig:thk_strong_coeff_expt} extends the testing range below $M_\infty=2$, at which conditions the DL-augmented Navier--Stokes, unmodified Navier--Stokes, DSMC, and experimental data converge.

\subsection{Shock Profile Asymmetry}

A second derived quantity, which provides a better assessment of the overall shape, is the density asymmetry quotient
\begin{equation}
    Q_p = \frac{\int_{-\infty}^{0} \rho^{*} \,dx }{\int_{0}^{\infty} \left[ 1-\rho^{*} \right]  \,dx },
\end{equation}
where $\rho^{*}$ is the normalized density, the maximum gradient of which is centered at $x=0$. The asymmetry quotient measures the skewness of a shock density profile relative to its midpoint; a perfectly symmetric profile has $Q_p=1$.

\begin{figure}
  \centering
  \includegraphics[width=0.6\textwidth]{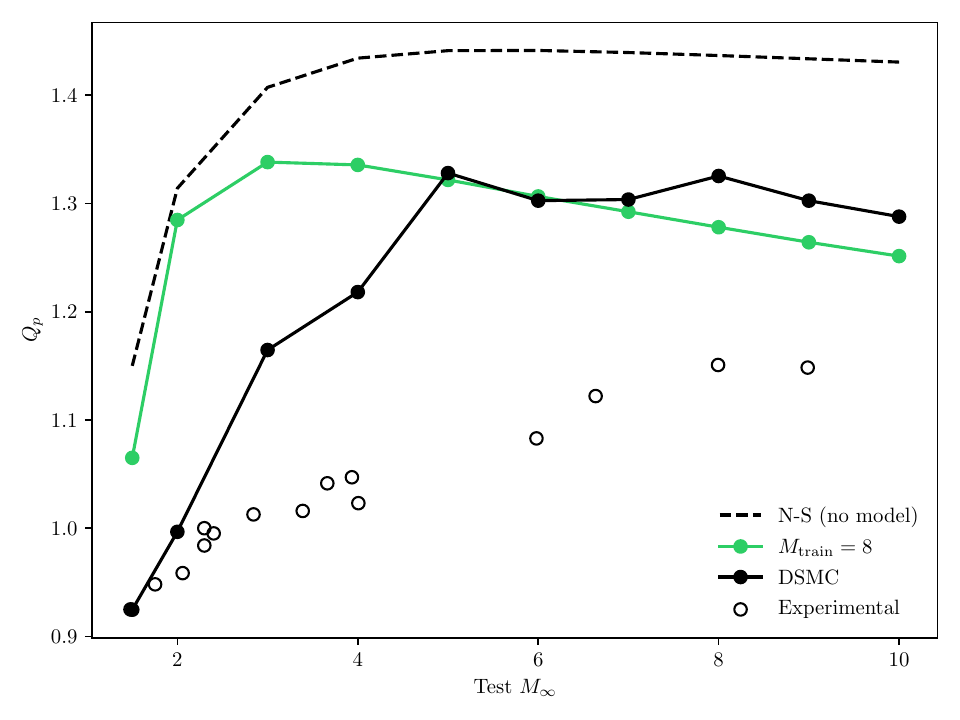}
  \caption{ Predicted density asymmetry quotient for the unmodified (no-model) Navier--Stokes equations, the augmented Navier--Stokes equations with the Approach A closure trained for $M_\mathrm{train}=8$, the DSMC target data, and experimental data~\cite{Alsemeyer1976}. }
  \label{fig:Qp_coeff_expt}
\end{figure}

Figure~\ref{fig:Qp_coeff_expt} compares the computed $Q_p$ using the unmodified Navier--Stokes equations, the Approach A model trained for $\Mtr=8$, the DSMC target data, and published experimental data~\cite{Alsemeyer1976}. The experimental data shows downstream-skewed shocks ($Q_p<1$) for freestream Mach numbers less than 2.5 and upstream-skewed shocks ($Q_p>1$) for higher Mach numbers, while the unmodified Navier--Stokes equations predict upstream-skewed shocks for all Mach numbers. The DSMC target profiles capture the overall experimental trends but lie between the experimental and unmodified Navier--Stokes profiles; this is consistent with other DSMC results using the VHS model \cite{torrilhon2004regularized,bentley2009investigation, bentley2009using, bentley2012shock}. The DL model leads to skewnesses reasonably close to the target DSMC profiles near the training $\Mtr=8$ condition, though these are still qualitatively different from the experimental measurements at higher Mach numbers.

\subsection{Training Convergence}

\begin{figure}
  \centering
  \includegraphics[width=0.49\textwidth]{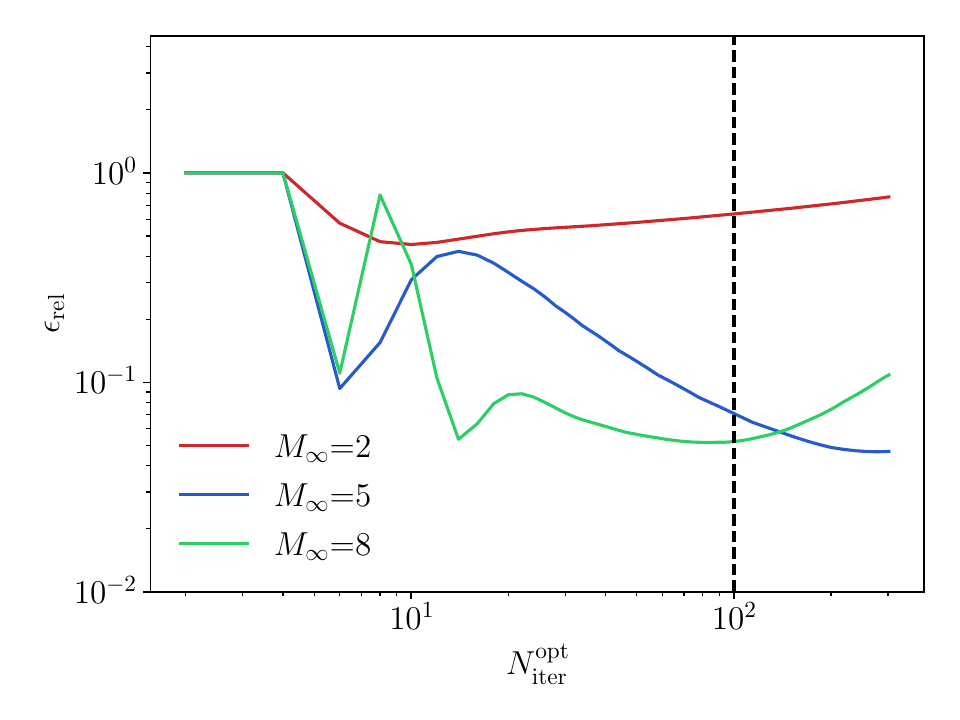}
  \includegraphics[width=0.49\textwidth]{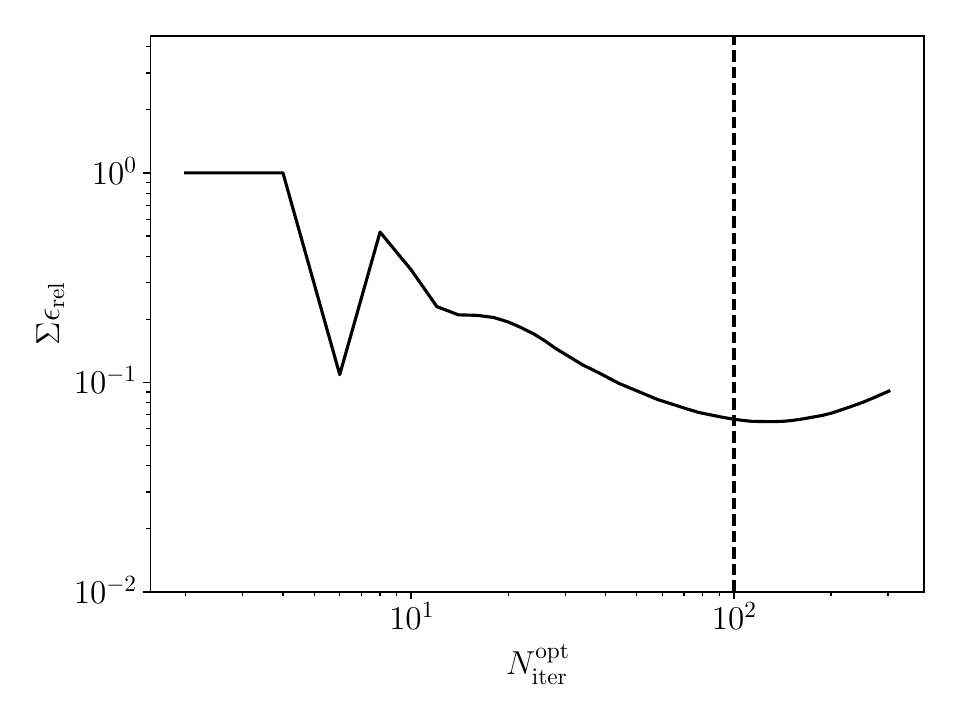}
  \caption{Convergence of $\epsilon_{\mathbf{rel}}$ (left) and the sum of $\epsilon_{\mathbf{rel}}$ for the three training Mach numbers (right) as a function of the number of optimization iterations $N_\mathrm{iter}^\mathrm{opt}$ using Approach A, for the training regimen $\Mtr=(2,5,8)$. The trained model at $N_\mathrm{iter}^\mathrm{opt} = 100$ ({\protect\dashedrule}) is used for in- and out-of-sample prediction.  }
  \label{fig:loss_all}
\end{figure}

The convergence of $\relErr$ with the number of optimization iterations is shown in Fig.~\ref{fig:loss_all} for the Approach A model trained simultaneously for $\Mtr=(2,5,8)$. A model trained for $N_\mathrm{iter}^\mathrm{opt} = 100$ iterations was used for the preceding tests, as it had the lowest total $\relErr$ between the three training Mach numbers. It can be seen that, as the model was trained for more iterations, its predictions became more accurate for the intermediate $M_{\infty}=5$ and less accurate for $M_{\infty}=2$ and 8. The testing errors of this model for all testing Mach numbers are listed in Table~\ref{tab:loss_Nopt_comp} for $N_\mathrm{iter}^\mathrm{opt} = 100$ and $N_\mathrm{iter}^\mathrm{opt} = 300$. It is evident that the prediction error for the longer-trained model is significantly higher for Mach numbers away from $M_{\infty}=5$, while interpolation to Mach numbers close to it are marginally lower. This is possibly evidence of the model overfitting to $M_{\infty}=5$. 

\begin{table}
  \centering
  \begin{tabular}{c c c c c c c c c c }
    \toprule
    & \multicolumn{9}{c}{Testing $M_{\infty}$} \\
    & 2 & 3 & 4 & 5 & 6 & 7 & 8 & 9 & 10   \\
    \midrule
    $N_\mathrm{iter}^\mathrm{opt} = 100$ & 0.026 & 0.071 & 0.117 & 0.148 & 0.156 & 0.199 & 0.199 & 0.205 & 0.250   \\
    $N_\mathrm{iter}^\mathrm{opt} = 300$ & 0.037 & 0.066 & 0.115 & 0.137 & 0.135 & 0.180 & 0.231 & 0.266 & 0.297   \\
    \bottomrule
  \end{tabular}
  \caption{Objective-function relative error $\epsilon_\mathrm{rel}$ for $N_\mathrm{iter}^\mathrm{opt} = 40$ and $N_\mathrm{iter}^\mathrm{opt} = 300$. Shown for the Approach A model trained for $\Mtr=(2,5,8)$.  }
  \label{tab:loss_Nopt_comp}
\end{table}

\subsection{Physical Interpretation of the Closure}
\begin{figure}
  \centering
  \includegraphics[width=0.49\textwidth]{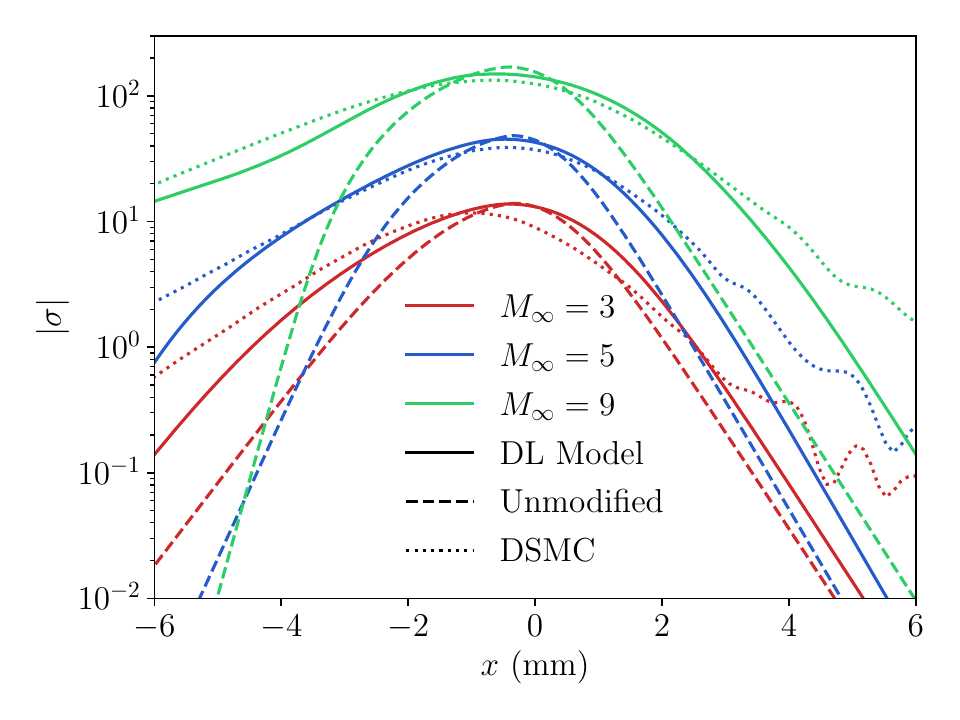}
  \includegraphics[width=0.49\textwidth]{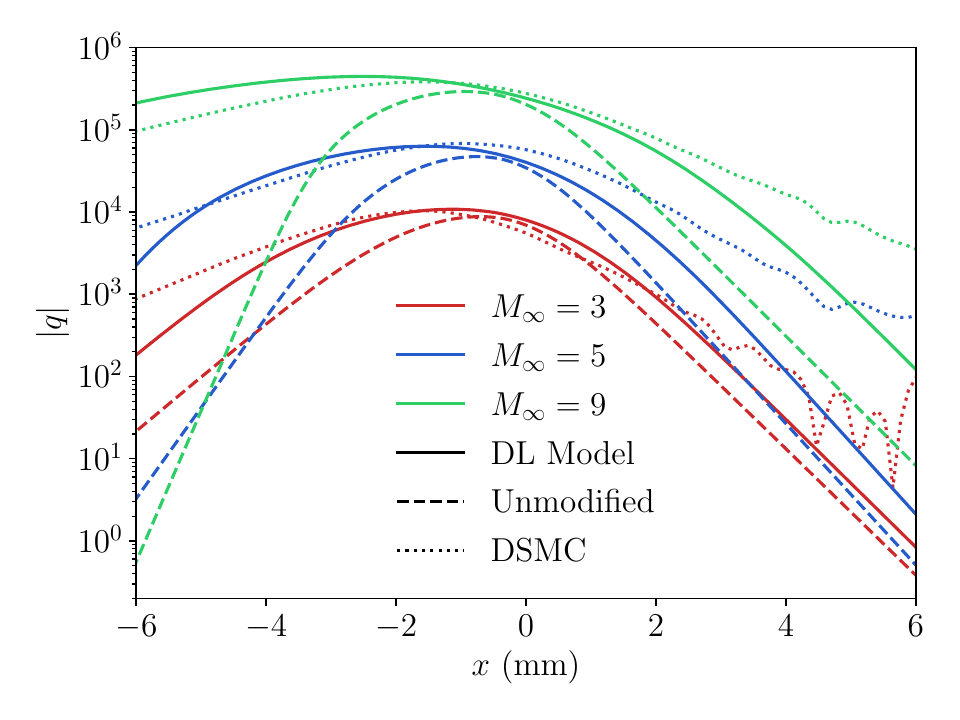}
  \caption{DL-modeled viscous stress and heat flux ({\protect\solidrule}) for Approach A versus the standard continuum models ({\protect\dashedrule}). Shown for the Approach A DL model trained for $\Mtr=8$. }
  \label{fig:model_form_coeff}
\end{figure}

\begin{figure}
  \centering
  \includegraphics[width=0.49\textwidth]{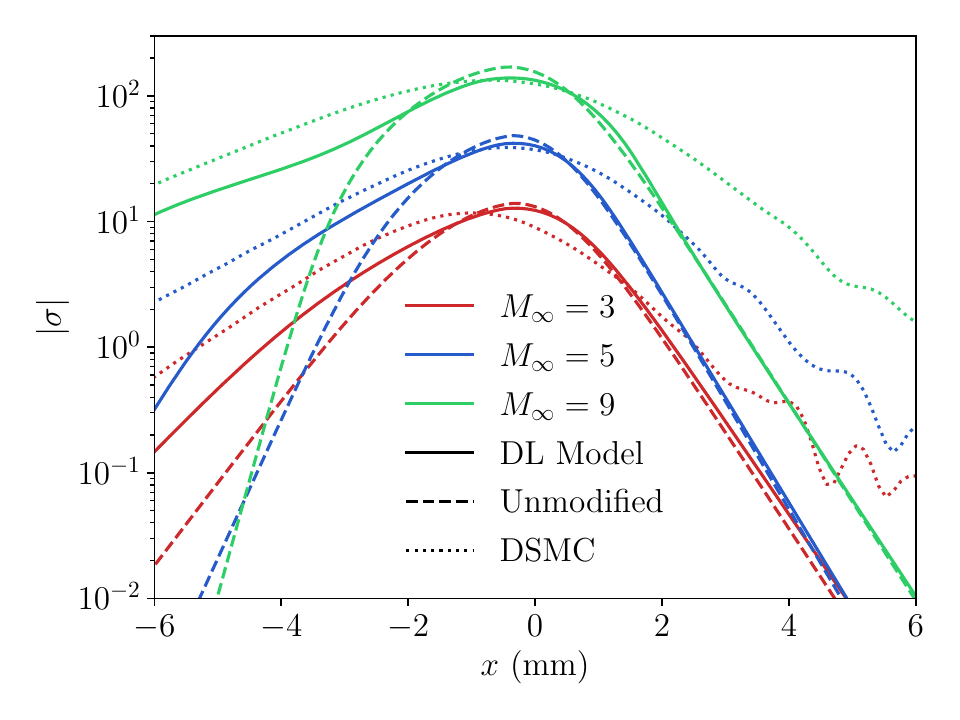}
  \includegraphics[width=0.49\textwidth]{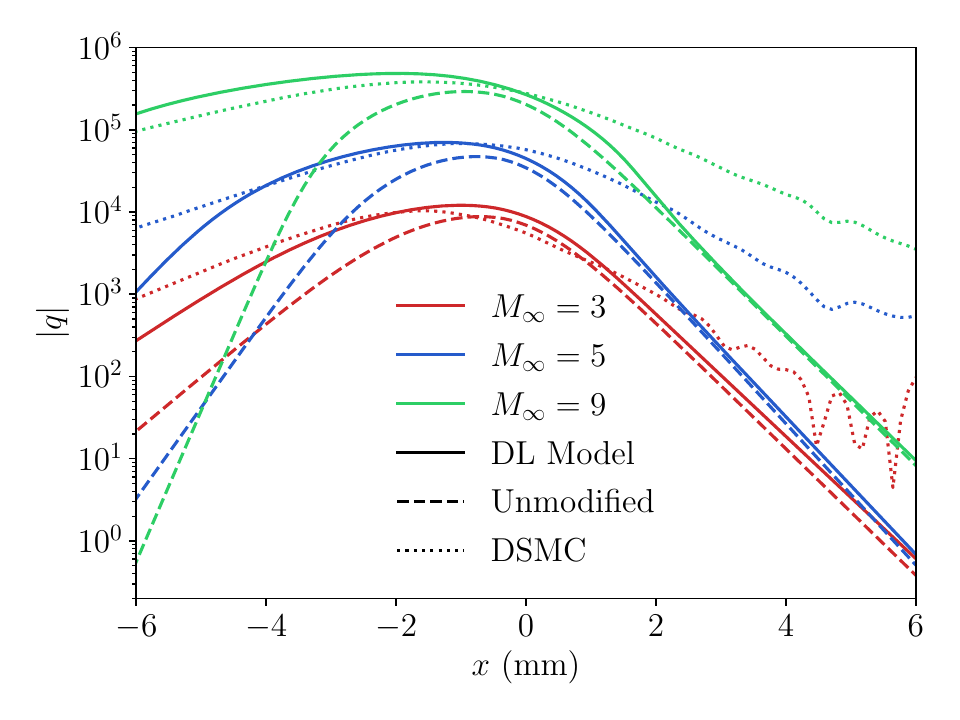}
  \caption{DL-modeled viscous stress and heat flux ({\protect\solidrule}) for Approach B versus the standard continuum models ({\protect\dashedrule}). Shown for the Approach B DL model trained for $\Mtr=8$. }
  \label{fig:model_form_strong}
\end{figure}

Figures~\ref{fig:model_form_coeff} and \ref{fig:model_form_strong} for Approaches A and B, respectively, compare the DL closure models to the unmodified continuum closures and the corresponding closures integrated from the DSMC data~\cite{Vincenti1917}. Approaches A and B modify the viscous stress and  heat flux to similar extents relative to the standard continuum models in the vicinity of the shock and are in excellent agreement with the DSMC profiles, particularly upstream and in the vicinity of the shock. Overall, the standard models underpredict molecular transport at these nonequilibrium conditions, especially the post-shock magnitude of the transport terms at higher freestream Mach numbers. Near the shock, the DL closure models and DSMC profiles have close magnitudes and are spatially more diffuse than the standard continuum models. This is consistent with the resulting steeper shock profiles (Fig.~\ref{fig:thk_strong_coeff_expt}).

Fig.~\ref{fig:model_form_coeff}, for Approach A, shows that the modified heat-flux and viscous-stress terms have the same slope away from the shock as the DSMC terms, unlike for Approach B. This is not surprising, as Approach B augments the continuum closures with nonlinear terms which, are functions of the (nearly zero) flow gradients away from the shock. Significantly, the DL-predicted profiles for $\sigma$ and $\bq$ differ, which reflects the fact that viscosity and thermal conductivity are continuum manifestations of different sub-continuum processes. Conversely, the temperature-dependent models have the same functional form \eqref{eq:mu_phys} with only slightly different coefficients. (Indeed, a common approximation for $\lambda(T)$ is based upon multiplying $\mu(T)$ by a constant Prandtl number.) This modeling choice renders the standard models unable to account for the differing degrees of momentum and thermal diffusion at transition-continuum conditions. The Approach A and B models have comparable maxima to the DSMC profiles, but the Approach B models generally underpredict $\sigma$ and $\bq$ downstream of the shock, which is consistent with their relatively inaccurate shock-thickness predictions. Furthermore, for both Approaches A and B, the modeled $\tau$ and $\bq$ are more asymmetric (upstream-skewed) than the DSMC, particularly for high Mach numbers, which leads to asymmetry quotients $Q_p > 1$ (see Fig.~\ref{fig:Qp_coeff_expt}).


\begin{figure}
  \centering
  \includegraphics[width=0.49\textwidth]{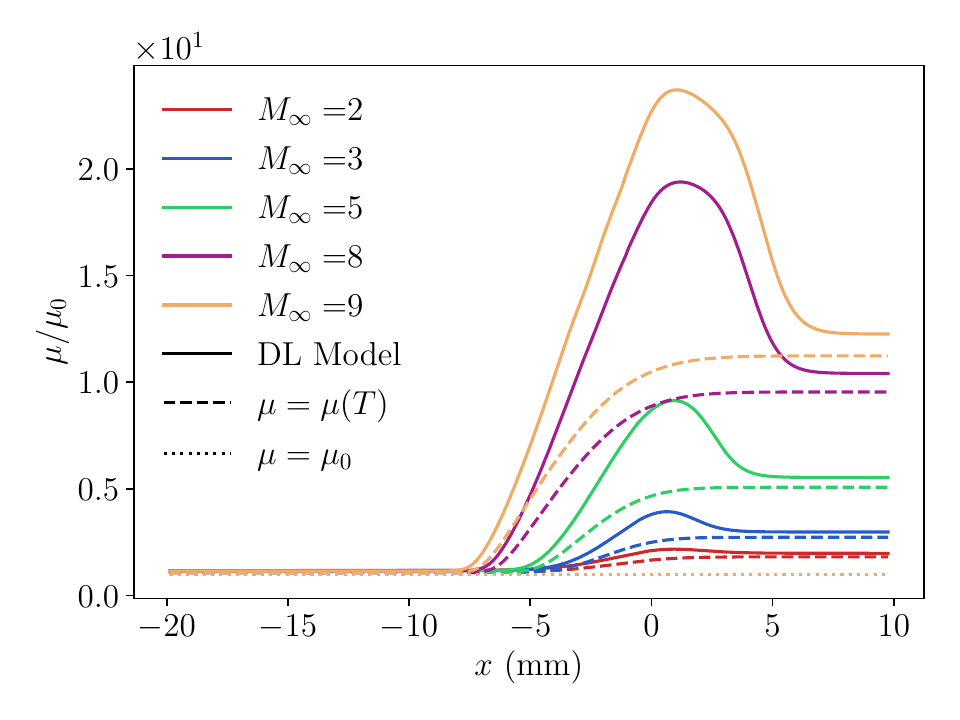}
  \includegraphics[width=0.49\textwidth]{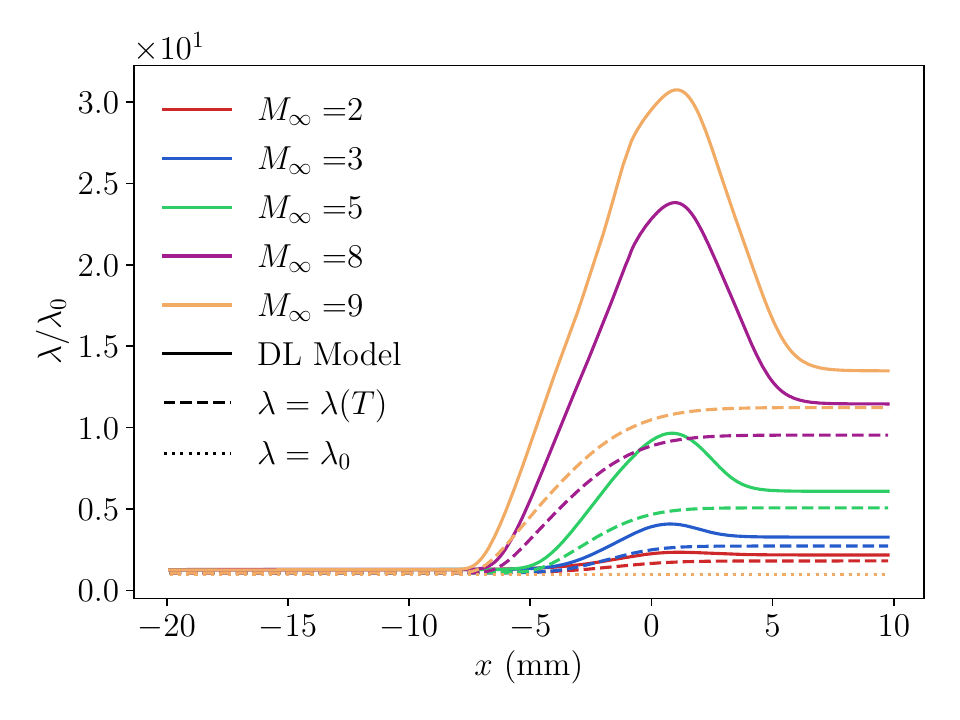}
  \caption{DL-modeled viscosity and thermal conductivity ({\protect\solidrule}; Eq.~\ref{eq:mu_aug_model}) versus standard temperature-dependent models ({\protect\dashedrule}; Eq.~\ref{eq:mu_phys}). Shown for the Approach A \eqref{eq:mu_aug_model_closure} DL model trained for $\Mtr=8$. }
  \label{fig:model_form}
\end{figure}

Approach A \eqref{eq:mu_aug_model_closure} modifies the viscosity and thermal conductivity; it is therefore instructive to consider the form of the model.
Fig.~\ref{fig:model_form} compares the Approach A closures for $\mu(\bU;\theta)$ and $\lambda(\bU;\theta)$ \eqref{eq:mu_aug_model}, evaluated using the neural network trained for $\Mtr=(2,5,8)$, to the standard, temperature-dependent models \eqref{eq:mu_phys} for the range of in- and out-of-sample testing Mach numbers.
The differences between the two models are pronounced at higher $M_\infty$ and both models predict significantly higher maximum values than the temperature-dependent models. It is also evident that downstream of the shock, the DL model predicted transport coefficients almost exactly match temperature dependent models, thus it accurately recovers continuum behavior away from the shock.

The general trends captured by the DL models could be incorporated into algebraic models similar to \eqref{eq:mu_phys}, which would enable computationally efficient predictions on a wide range of out-of-sample Mach numbers and geometric configurations, even with higher dimensionality than the present shock configurations. This is more probable for simpler models, such as our Approach A, that interact only with a scalar coefficient, and so could be more portable to different dimensionalities, flow geometries, and/or alignments of key features (e.g., shocks) with the computational mesh. The more complicated models that introduce vector/tensor modifications, such as Approach B, are more likely to train in a geometry-specific manner and likely require re-training for new geometries. The development and validity of more generally applicable models, as well as the extrapolability of trained models to different freestream thermodynamic conditions and gas compositions, will be the subject of future work.

\section{Conclusion} \label{sec:conclusion}

This paper presents an adjoint-based deep learning augmentation method to extend the validity of the continuum flow equations to the transition-continuum regime. Two modeling approaches are compared: Approach A modifies only the viscosity and thermal conductivity, while
Approach B directly augments the continuum diffusive fluxes with corrective neural networks. These modeling approaches have less flexibility than outright replacement of the fluxes but were more stable during optimization. Model training is performed by optimizing over the Navier--Stokes equations while targeting trusted flow data from DSMC solutions of the Boltzmann equation.

An online adjoint optimization method enables this PDE-constrained optimization to be performed efficiently.  The trained models improve the accuracy of Navier--Stokes predictions in this regime and exhibit stability and extrapolability for Mach numbers higher and lower than those used for training, with concomitant reduction of accuracy when used far out-of-sample (although still not of lower accuracy than that of the unmodified Navier--Stokes equations). Models trained using Approach A have marginally better accuracy compared to models trained using Approach B, and part of Approach A's success may be attributed to the distinctions it identifies between the different transport-coefficient profiles and their different growth trends with respect to the upstream Mach number. This results in different profiles for the DL-predicted viscous-stress and heat-flux closures. 

Entropy constraints  ensure that the learned closure models satisfy the second law of thermodynamics. The strong entropy constraint, which poses an algebraic constraint on the output of the DL model, resulted in stable, accurate predictions over the range of testing Mach numbers. The weak entropy constraint, which penalizes the violation of the entropy inequality in the optimization loss function, is attractive for its flexibility but was less stable extrapolating to higher Mach numbers. Further study of the weighting factor for the entropy violation term is necessary to realize the full potential of the weak entropy condition.

Notably, the loss function for the online adjoint optimization can be arbitrarily defined, can include flow variables (the PDE solution or derived quantities), and does not need to explicitly include the unclosed term(s). This is in contrast to \emph{a priori} training, which conforms the model directly to the unclosed terms.
The success of the online adjoint method with ``third-party'' DSMC target data (i.e., not directly obtained from the flow PDEs) is striking, though not unexpected, given the success of adjoint-based methods for aerodynamic shape design and turbulence modeling, among others. One hope is that this method will enable model development against experimental target data with comparable stability and accuracy, though this remains to be tested. Alternatively, or in addition, the loss function for the flow model can target derived or integral quantities that are functions of the flow variables, such as the shock thickness or, for wall-bounded flows, the boundary layer thickness and drag coefficient.

Future work will focus on extending the current model to more-complex geometric scenarios (2D oblique and curved shocks, 3D shocks) and flows with additional physical complexity (nonmonatomic gases, reacting flows). Particularly useful for cases with additional physical complexity, the adjoint-based DL approach  can account for multiple, coupled constitutive models simultaneously, which would enable, for example, coupling the transport coefficients with the chemical kinetics.

\section*{Acknowledgments}
The authors would like to thank Prof.~Iain Boyd for sharing his DSMC solver. This work was supported by the Office of Naval Research under award N00014-22-1-2441. The research of M.P.\ is supported by the Vannevar Bush Faculty Fellowship. The authors acknowledge computational time on resources supported by the University of Notre Dame Center for Research Computing (CRC).

\bibliography{library}

\end{document}